\documentclass[conference]{IEEEtran}
\usepackage{url}
\usepackage{lipsum}
\usepackage{todonotes}
\usepackage{amsmath}
\usepackage{graphicx}
\usepackage{paralist}
\usepackage{comment}
\usepackage{capt-of}
\usepackage{booktabs}
\usepackage[caption=true]{subfig}
\usepackage{siunitx}
\usepackage{multirow}
\usepackage{mdframed}
\usepackage{tabularx}

\pdfoutput=1
\begin{document}

\title{Practical Hidden Voice Attacks against Speech and Speaker Recognition Systems}
\author{\IEEEauthorblockN{Hadi Abdullah, Washington Garcia, Christian Peeters, Patrick Traynor, Kevin  R. B. Butler and Joseph Wilson}
\IEEEauthorblockA{University of Florida\\
\{hadi10102, w.garcia, cpeeters, traynor, butler, jnw\}@ufl.edu
}
}
\IEEEoverridecommandlockouts
\makeatletter\def\@IEEEpubidpullup{6.5\baselineskip}\makeatother

\maketitle
\pagestyle{plain}
\begin{abstract}

Voice Processing Systems (VPSes), now widely deployed, have been made significantly more accurate
through the application of recent advances in machine learning.
However, adversarial machine learning has similarly advanced and has been used to
demonstrate that VPSes are vulnerable to the injection of hidden
commands - audio obscured by noise that is correctly recognized by a VPS
but not by human beings. Such attacks, though, are often highly
dependent on white-box knowledge of a specific machine learning model
and limited to specific microphones and speakers, making
their use across different acoustic hardware platforms (and
thus their practicality) limited. In this paper, we break these dependencies and
make hidden command attacks more practical through model-agnostic
(black-box) attacks, which exploit knowledge of the signal processing
algorithms commonly used by VPSes to generate the data fed into machine
learning systems.  Specifically, we exploit the fact that multiple
source audio samples have similar feature vectors when transformed by
acoustic feature extraction algorithms (e.g., FFTs). We develop four
classes of perturbations that create unintelligible audio and test them
against 12 machine learning models, including 7 proprietary models
(e.g., Google Speech API, Bing Speech API, IBM Speech API, Azure Speaker
API, etc), and demonstrate successful attacks against all targets.
Moreover, we successfully use our maliciously generated audio samples in
multiple hardware configurations, demonstrating effectiveness across
both models and real systems. In so doing, we demonstrate that
domain-specific knowledge of audio signal processing represents a
practical means of generating successful hidden voice command attacks.

\end{abstract}

\section{Introduction}
\label{sec:intro}

Voice Processing Systems (VPSes) are rapidly becoming the primary means
by which users interface with devices. In particular, the increasing use
of constrained/headless devices (e.g., mobile phones, digital home
assistants) has led to their widespread deployment. These
interfaces have been widely heralded not only for simplifying
interaction for traditional users, but also for dramatically expanding
inclusion for disabled communities and the elderly~\cite{CM17, HS18}.

The driving force behind practical voice-driven systems has been
foundational advances in machine learning. Models incorporating
seemingly ever-increasing complexity now handle massive quantities of
data with ease. When combined with well-known techniques from signal
processing for feature extraction, such systems now provide highly
accurate speech and speaker recognition. However, VPSes also introduce
substantial security problems. As has been demonstrated
intentionally~\cite{BKAd} and unintentionally~\cite{UKdollhouse}, these
interfaces often recognize and execute commands from any nearby device
capable of playing audio. Moreover, recent research has demonstrated
that attackers with white-box knowledge of machine learning models can
generate audio samples that are correctly transcribed by VPSes but
difficult for humans to understand~\cite{carlini2016hidden,
yuan2018commandersong}. 

This work takes a different approach. Instead of attacking specific
machine learning models, we instead take advantage of the \textit{signal
processing} phase of VPSes. In particular, because nearly all speech and
speaker recognition models appear to rely on a finite set of features
from classical signal processing (e.g., frequencies from FFTs,
coefficients from MFCs), we demonstrate that modifying audio to
produce similar feature vectors allows us to perform powerful attacks against
machine learning systems in a black-box fashion.

In so doing, we make the following contributions:

\begin{figure*} 
  \centering
    \includegraphics[width=1.0\linewidth]{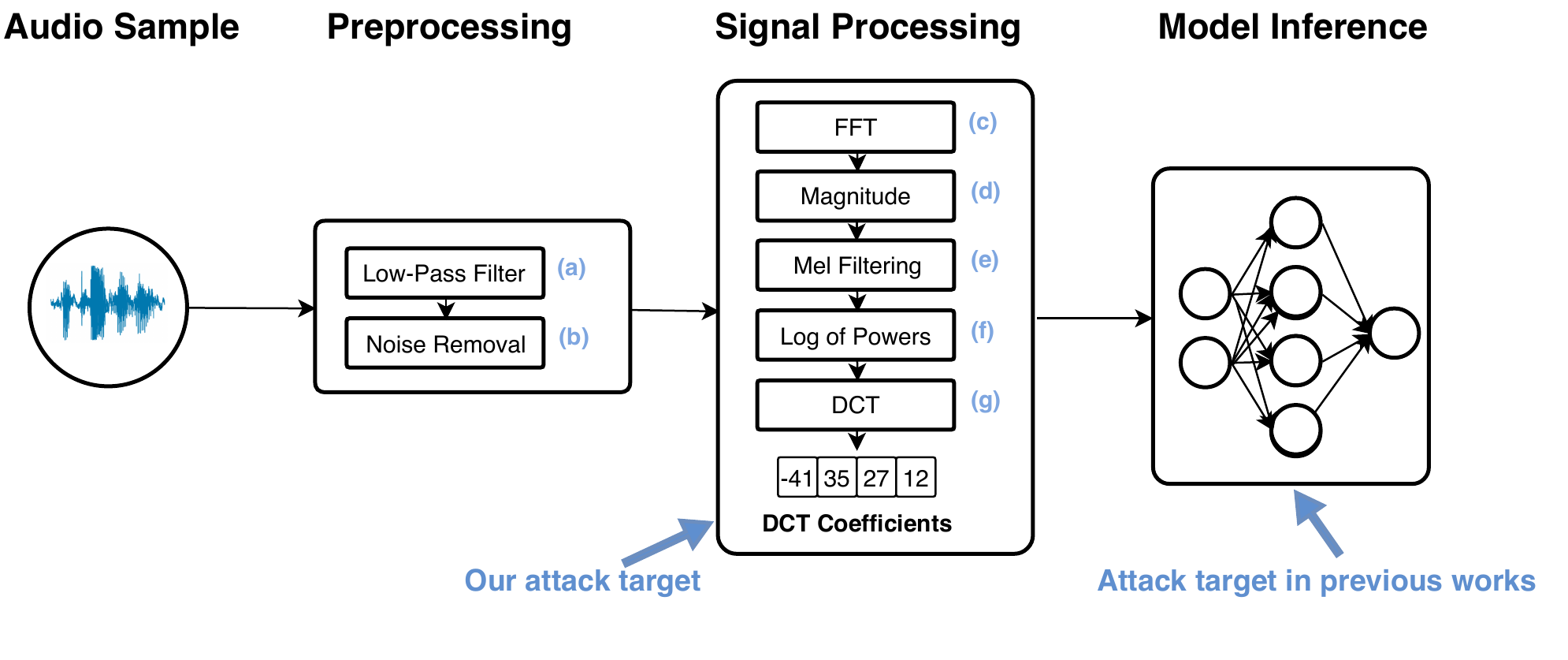}
    \caption{This generic VPS workflow for audio transcription illustrates the various processing steps that are carried out on the audio before it is passed to the Machine Learning Model for transcription. First, the audio is preprocessed to remove noise. Next, the audio is passed through a signal processing algorithm that changes the format of the input, retains the important features and discards the rest. Lastly, these features are given to a machine learning model for inference. }
  \vspace{-1em}
\label{fig:ML_workflow}
\end{figure*}

\begin{itemize}

    \item {\bf Develop perturbations during the signal processing phase:} The
key insight of this work is that many pieces of source audio can be
transformed into the same feature vector used by machine learning models.
That is, by attacking the feature extraction of the signal processing
phase as opposed to the model itself, we are able to generate over
20,000 audio samples that pass for meaningful audio while sounding like
unintelligible noise.  Accordingly, VPSes can be attacked quickly
(i.e., in a matter of seconds) and effectively in an entirely black-box fashion.

    \item {\bf Demonstrate attack and hardware independence:} Previous
research has been highly dependent on knowledge of the model being used
and constrained to specific pieces of hardware. For example, Yuan et al.
evaluated one white-box and two black-box
models~\cite{yuan2018commandersong}, while Carlini et al. evaluated one
black-box and one white-box model~\cite{carlini2016hidden}. We evaluate
our attacks against 12 black-box models that use a variety of machine
learning algorithms, including seven proprietary and five local
models; however, in no case do we use any information about the model or
weights (i.e., we treat all models, even if publicly available, as black
boxes). We demonstrate successful attacks using multiple sets of
speakers and microphones. Our attack evaluation is far more
comprehensive than prior work in adversarial audio. Such wide
effectiveness of attacks has not previously been demonstrated.

    \item {\bf Use psychoacoustics to worsen audio intelligibility:}
Prior work has focused on adding background noise to obfuscate commands;
however, such noise more often than not reduces successful
transcription.  We take advantage of the fact that humans have
difficulties interpreting speech in the presence of certain classes of
noise {\it within the range of human hearing} (unlike the Dolphin Attack
which injects ultrasonic audio~\cite{zhang2017dolphinattack} and can
therefore easily be filtered out).

\end{itemize}

We believe that by demonstrating the above contributions we take hidden
voice commands from the realm of {\it possible to practical}.  Moreover,
it is our belief that such commands are not easily countered by making
machine learning models more robust, because the feature vectors that the
models receive are virtually identical whether the VPS is being attacked
or not. Accordingly, it is crucial that system designers consider
adversaries with significant expertise in the domain in which their
machine learning model operates.


The remainder of the paper is organized as follows:
Section~\ref{sec:background} provides background information on signal
processing, speech and speaker recognition, and machine learning;
Section~\ref{sec:hypothesis} presents our hypothesis;
Section~\ref{sec:methodology} discusses our threat model, our
methodology for generating perturbations, and experimental setup;
Section~\ref{sec:results} shows our experimental results against a range
of speech and speaker recognition models; Section~\ref{sec:discussion}
discusses psychoacoustics and potential defenses;
Section~\ref{sec:relwork} presents related work; and
Section~\ref{sec:conclusion} offers concluding remarks.

\section{Background}\label{sec:background}

\subsection{Voice Processing System (VPS)}
Any machine learning based voice processing tool can be considered a VPS. In this paper, we use the term to refer to both Automatic Speech Recognition (ASR) and Speaker Identification models.

\subsubsection{ASRs} An ASR converts raw human audio to text. Generally, most ASRs accomplish this task using the steps shown in Figure~\ref{fig:ML_workflow}: pre-processing, signal processing and model inference. 

Preprocessing involves applying filters to the audio in order to remove background noise and any frequencies that are outside the range of the human audio tract, (Figure~\ref{fig:ML_workflow}a and \ref{fig:ML_workflow}b). Signal processing algorithms are then used to capture the most important features and characteristics, reducing the dimensionality of the audio. The signal processing step outputs a feature vector. Most ASRs employ the Mel-Frequency Cepstrum Coefficient (MFCC) algorithm, for feature extraction, because of its ability to extrapolate important features, similar to the human ear.  The feature vector is then passed to the model for either training or inferencing.

\subsubsection{Speaker Identification model} Speaker Identification models identify the speaker in a recording by comparing voice samples of the speakers the model was trained on. The internal workings of the Speaker Identification model are largely similar to that of ASRs, with an additional voting scheme. For each audio subsample, the model assigns a vote for the speaker the subsample most likely belongs to. After processing the entire audio file, the votes are tallied. The speaker with the largest number of votes is designated as the source of the input audio sample.

\subsection{Signal Processing}
Signal processing is a major components of all VPSes. These algorithms capture only the most important aspects of the data. The ability of the signal processing algorithm to properly identify the important aspects of the data is directly related to the quality of training a machine learning model undergoes.

\subsubsection{Mel-Frequency Cepstrum Coefficient (MFCC)}
The method for obtaining an MFCC vector of an audio sample begins by first breaking up the audio into 20 ms windows. Each window goes through four major steps as seen in Figure~\ref{fig:ML_workflow}.

\noindent\textbf{Fast Fourier Transform (FFT) and Magnitude:}
For each window, an FFT and its magnitude are taken which generates a frequency domain representation of the audio (Figure~\ref{fig:ML_workflow}c and Figure~\ref{fig:ML_workflow}d) called a magnitude spectrum. The magnitude spectrum details each frequency and the corresponding intensity that make up a signal.

\noindent\textbf{Mel Filtering:}
The Mel scale translates actual differences in frequencies to perceived differences in frequencies by the human ear. Frequency data is mapped to the Mel scale (Figure~\ref{fig:ML_workflow}e) using triangular overlapping windows, known as Mel filter banks. 

\noindent\textbf{Logarithm of Powers:}
To mimic the way in which human hearing perceives loudness, the energies of each Mel filter bank are then put on a logarithmic scale (Figure~\ref{fig:ML_workflow}f).

\noindent\textbf{Discrete Cosine Transform (DCT):}
The final step in obtaining the MFCCs is to take the discrete cosine transform of the list of Mel filter bank energies (Figure~\ref{fig:ML_workflow}g). The result is a vector of the MFCCs.

\subsubsection{Other Methods}

There are a variety of other signal processing techniques used in modern
VPSes as well. Examples of these include Mel-Frequency Spectral
Coefficients (MFSC), Linear Predictive Coding, and Perceptual Linear
Prediction (PLP)~\cite{rabiner1978digital}. Much like MFCCs, these
represent deterministic techniques for signal processing. Other VPSes employ
probabilistic techniques, specifically transfer
learning~\cite{torrey2010transfer}. In this case, one model is trained
to learn how to extract features from the input, while the other model
is fed these features for inferencing. This layered approach to VPSes is
a recent development~\cite{venugopalan2014translating}. Additionally, some
VPSes, called ``end-to-end" systems, replace all
intermediate modules between the raw input and the
model by removing pre-processing and signal processing steps~\cite{graves2014towards}~\cite{hannun2014deep}. These systems
aim to remove the increased engineering effort required in bootstrapping
additional modules, as well as increased processing time and complicated
system deployment~\cite{amodei2016deep}.

\subsection{Model Inference}
Features gathered from the signal processing steps are then passed onto the machine learning algorithms for inference. VPSes make extensive use of machine learning algorithms for speech recognition. However, the definition of such systems varies due to the evolving nature of machine learning techniques. In general, a machine learning system can be described as a learned mapping between a set of inputs $X \in {\rm I\!R}^D$, for some input dimensionality $D$, to a set of outputs $Y$ that consist of either continuous or categorical values, such that $X \rightarrow Y$. To learn such a mapping, the values are converted to the form $F(x) = y + \epsilon$, where $\epsilon$ is an error term to be minimized for all inputs $x \in X$ and known outputs $y \in Y$ of a piece of data known as the training set. A system is said to \textit{interpolate} between inputs and outputs of a training set to model the mapping $X \rightarrow Y$. Once the system is trained, it must then \textit{extrapolate} the knowledge it learned onto a new, unseen space of inputs, known as the test set. Rate of error on the test set is the most common measure of machine learning systems. In general, any input $X$ to the machine learning system is known as a \textit{feature}, as it constitutes a unique trait observable across all input samples, and ideally, can be used to distinguish between different samples.

Early machine learning systems specialized in specific areas and were designed to interpolate knowledge of some particular domain, and thus became \textit{domain experts}~\cite{Angluin1992}. Due to the specificity of such systems, they required extensive feature engineering, often involving human domain knowledge in order to produce useful results. Thus, much of the early work in this field revolves around feature design. With the advent of  ubiquitous methods of data collection, later methods focused on data-driven approaches, opting for more general knowledge rather than domain-specific inference. Modern systems are now capable of automatically performing what was previously known as feature engineering, instead opting for \textit{feature extraction} from a large, unstructured corpus of data. Thus, these techniques can automatically learn the correct features relevant to create the mapping and apply a pre-defined cost function to minimize the error term. Due to the absence of specific feature engineering, these systems are not only more flexible, but can be split into independent modules to be used across different domains or applications (i.e., transfer learning), and also offer much better extrapolation performance. 
\begin{figure*} 
  \centering
    \includegraphics[width=1.0\textwidth]{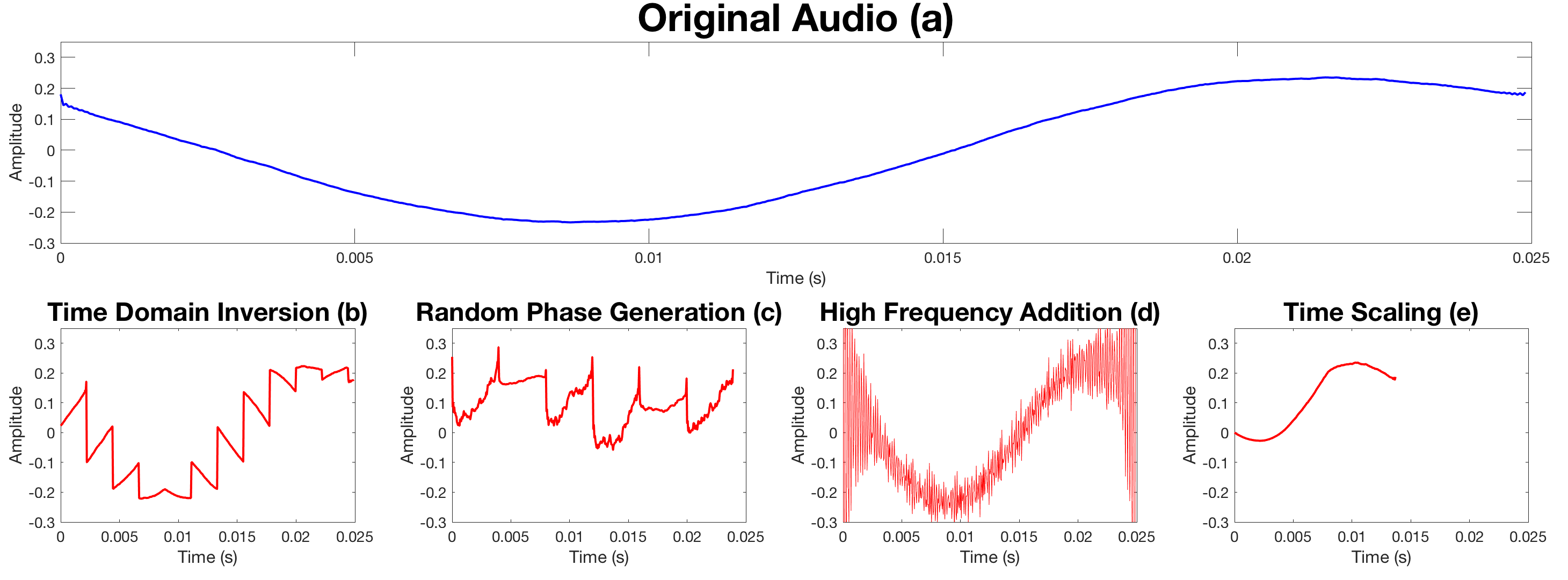}
  \caption{The above figure shows the different perturbation techniques applied to the original signal (a). Signals (b) to (e) show the result of applying the perturbation schemes to (a).}
\label{fig:perturbations}
\vspace{-1em}
\end{figure*}

\subsection{Psychoacoustics}
Psychoacoustics is the science of how humans perceive audio. There is a broad and deep field of research dedicated to this topic that continues to advance our understanding of human hearing and the way the brain interprets acoustic signals. Research in this field has found many subtleties in the way people hear and interpret speech. Accordingly, successful hidden commands must consider psychoacoustic effects. 

Studies in psychoacoustics have demonstrated that human hearing and perception of sound has areas of both robustness and weakness. When in the presence of multiple sources of sound, it is easy for humans to focus on a single source. This is known as the Cocktail Party effect~\cite{cherry}, where humans can tune out sound that is not particularly pertinent. Human hearing is also robust at interpreting speech in the presence of added background noise~\cite{noise}. This means that simply adding noise to recorded speech will not decrease its intelligibility, only suppress the volume of the speech.

Alternatively, human hearing and perception can also be weak, particularly with higher frequencies. Humans hear sound on a logarithmic scale and tend to perceive higher frequencies as louder~\cite{ISO226}. The Mel scale~\cite{mel_scale}, as well as the concept of similar frequency masking~\cite{masking_book}, show that people poorly discern differences in high frequencies compared to low frequencies. Humans are also poor at interpreting discontinuous or random sound, whereas they tend to enjoy smooth and continuous audio signals. However, discontinuous and more erratic signals are, by definition, noise~\cite{method_noise}. In addition to being unintelligible, these signals tend to be jarring and discomforting to most individuals.  

We can take advantage of psychoacoustics to better perturb voice commands. We discuss our use of psychoacoustics more in Section~\ref{sec:discussion}.

\section{Hypothesis}
\label{sec:hypothesis}

Human speech provides a limited set of features for training and testing
ASR models: most aspects of speech are filtered out during
preprocessing or feature extraction algorithms. This reduced subset of
the original input is passed as a feature vector onto the model for training or testing.
This provides opportunities to introduce acoustic artifacts, via
perturbation, to the audio. Such artifacts are removed during the
signal processing steps.  By using knowledge about psychoacoustics and the physical characteristics of audio, we can develop
attack audio that is agnostic of any particular machine learning model. While these
artifacts will make human interpretability of the audio difficult, they
will have no impact on the ASR's recognition ability, leaving ASRs exposed to hidden voice command
attacks.

\section{Methodology}
\label{sec:methodology}

We develop perturbations to generate attack audio that is
uninterpretable by the user, but correctly inferred by the VPS. This
section describes the attack scenario, the capabilities required to
conduct a successful attack, perturbation schemes that generate attack
audio samples, and the VPSes we tested the attacks against.

\subsection{Attack Scenario}

An adversary wants to execute an unauthorized command on a VPS (e.g.,
Amazon Alexa, Google Home Assistant, or any voice activated device). To
do so, the adversary plays an obfuscated audio command in the direction
of the VPS. The target VPS uses an ASR internally to transcribe the
audio commands in order to execute them. In addition to the ASR, the VPS
can also use a speaker identification model to authenticate the speaker
before executing the command. The obfuscated audio could contain orders
for example to initiate a wire-transfer, unlock door, or generally cause
the execution of functionality available to a legitimate user. The
attack audio is played via a compromised IoT device or directional
speaker. Although the owner of the target VPS may be within audible
range, they will not recognize the attack audio as a voice command.
Hence, the owner will remain unaware that an attack is in progress.

An attack against these systems will occur based on the
standard exploit development, probe, and attack strategy. That is, we
expect an attacker to develop a corpus of attack audio that successfully
activates one or more models. They will then probe a potential target,
learning which model they are targeting either through direct
information or via probing. The adversary will then select sample attack
audio from its corpus and execute the attack. 

\subsection{Threat Model}

We assume no knowledge of the model (i.e., black-box attack). The
attacker does not need to know the type or internal architecture of the
victim model (i.e., the number of layers or the internal weights).
However, the adversary is familiar with audio and speech processing. 

Our attack generates noise-like attack audio that will be effective
against VPSes: in transcription tasks, the perturbed audio is
transcribed the same as original unperturbed audio, while in
identification tasks, the system will identify the perturbed audio as
the voice of the original speaker. The perturbation methods are designed
to exploit assumptions that VPSes make about acoustic properties.

In transcription tasks, the adversary has a sample of correctly
transcribed audio. Similarly, for attacking a speaker model, the
adversary has a voice sample of the victim speaker that the target VPS
will correctly identify. The perturbation methods are designed to
maintain the normal audio sample's important acoustic properties while
altering its audible perception. If the audio sample is incomprehensible
to the target model without the perturbation, it will remain so with the
perturbation. 
\textcolor{black}{ The threat model is designed to emulate that of previous attack papers, specifically Carlini et al~\cite{carlini2016hidden}. The attacker is not located in the room, but is able to use the speaker remotely. The victim is in close vicinity of the attack speakers but is not actively listening for or expecting an attack. The victim might hear the attack audio, but is unable to decipher it, thus would not know that an attack is in progress.}

\textcolor{black}{The attacker has no knowledge of the acoustic hardware used by the victim. We assume that target ASR is close to the attack speaker. Homes are being equipped with an increasing number of speaker enabled IoT devices. This number is expected to rise in order for users to be able to continuously interact with the home assistant during their daily routine throughout the house. These speakers can be exploited remotely and then used by the attacker to play the attack audio. Additionally, the victim ASR devices, like Alexa, have an array of high quality microphones that can detect audio from a wide variety of locations and angles. In order to accommodate for the above factors, we assume that the target ASR is one foot away from the attack speaker.}

\subsection{Types of Perturbations}\label{sec:types_perturb}

We propose four perturbation techniques. Each resulting attack sample
includes one or more perturbations applied in succession which can be
used against the target VPS. We use Figure~\ref{fig:perturbations} as an
exemplar source and show each perturbation detailed below.

\subsubsection{Time Domain Inversion (TDI)}

Most VPSes use FFTs (Figure~\ref{fig:ML_workflow}c) to decompose a
signal into its composite frequencies, called a spectrum. The FFT is a
many-to-one function. This means two completely different signals in the
time domain can have similar spectra. Our TDI
perturbation method exploits this property by modifying the audio in the
time domain while preserving its spectrum, by inverting the windowed
signal. As shown in Figure~\ref{fig:perturbations}b, inverting small
windows across the entire signal removes the smoothness. Due to the
principles of psychoacoustics, this perturbed audio is difficult to
understand as the human ear interprets any discontinuous signal as
noisy~\cite{method_noise}.

\subsubsection{Random Phase Generation (RPG)}

For each frequency in the spectrum, the FFT returns the value in complex
form $a_0+b_0i$, where $a_0$ and $b_0$ define the phase of a signal. To
get the intensity at each frequency, the magnitude
(Figure~\ref{fig:ML_workflow}d) of the complex spectrum is taken  to
yield a magnitude spectrum using the equation below:

\begin{align}
	magnitude_{original}=Y = \sqrt{a_0^2+b_0^2i}
\end{align}

Because the magnitude function is many-to-one, there are multiple values
of $a$ and $b$ that have the same $magnitude_{original}$. Informally,
two signals of different phases can have the same magnitude spectrum.
This second perturbation method picks two random numbers $a_n$ and $b_n$
such that:

\begin{align}
	magnitude_{original}=Y = \sqrt{a_n^2+b_n^2i}
\end{align}

This outputs a new signal with a different phase, yet with the same
magnitude spectrum as the original signal as shown in
Figure~\ref{fig:perturbations}c. This will introduce similar
discontinuities in the signal, which makes the perturbed audio harder to
interpret due to the fundamentals of psychoacoustics.

\subsubsection{High Frequency Addition (HFA)}

During preprocessing, frequencies beyond the range of the human voice
are removed from the audio using a low-pass filter
(Figure~\ref{fig:ML_workflow}a) in order to improve VPS accuracy. In
most cases, this cut-off point is at least 8000 Hz, for two reasons: the majority
of spoken content is below this level, \textcolor{black}{and speech is typically sampled at 16000 Hz}\footnote{\textcolor{black}{If the system does not have hardware low pass filters, the audio signal will alias.}}.
The third perturbation method
adds high frequencies to the audio that are filtered out during the
preprocessing stage. We create high frequency sine waves and add it to
the real audio (Figure~\ref{fig:perturbations}d). If the sine waves have
enough intensity, it has the potential to mask the underlying audio
command to the human ear. The resulting audio may also become
potentially painful to listen to as the human ear is sensitive to high
frequencies. The psychoacoustic reasoning behind this is further
discussed in Section~\ref{sec:discussion}.

\subsubsection{Time Scaling (TS)}

Speaker and speech recognition models need to account for the speed of
human speech. It is harder for humans to comprehend words spoken at a
faster rate, relative to the same words spoken at a slower
rate~\cite{hanley1949effect}. In the fourth and final perturbation, we
can accelerate the voice commands to a point where they are still able
to be properly transcribed. We do so by compressing the audio in the
time domain by discarding unnecessary samples and maintaining the same
sample rate. As a result, the audio is shorter in time, but retains the
same spectrum as the original.

We generate a set of audio files that contain the same voice command but
vary in the degree of the audio's speed
(Figure~\ref{fig:perturbations}e). We then run these files against all
of the speech recognition models we observe and record the file with the
highest speed that was still properly transcribed. Though applying this
perturbation by itself may not completely hinder the ability of a human
to comprehend the original voice command, applying this in conjunction
with the other perturbations makes interpreting the audio difficult.

\begin{table*}
\centering
\begin{tabular}{|l|l|l|l|l|l|} 
\hline

\textbf{Voice Processing System}       & \textbf{Model Type} & \textbf{Task}                               & \textbf{Feature Extraction} & \textbf{Phrase ID} & Online/Local  \\ 
\hline
\textbf{Azure Verification API}~\cite{azure_verify}             & Unknown             & Identification                                 & Unknown                     & D               & Online        \\ 
\hline
\textbf{Azure Attestation API}~\cite{azure_attest}             & Unknown             & Identification                                 & Unknown                     & A,B,C           & Online        \\ 
\hline
\textbf{Bing Speech API}~\cite{bing}            & Unknown             & Transcription                               & Unknown                     & E,F,G,H         & Online        \\ 
\hline
\textbf{Google Client  Speech API}~\cite{google}    & Unknown             & \textcolor[rgb]{0.2,0.2,0.2}{Transcription} & Unknown                     & E,F,G,H         & Online        \\ 
\hline
\textbf{Houndify Speech API}~\cite{houndify}          & Unknown             & \textcolor[rgb]{0.2,0.2,0.2}{Transcription} & Unknown                     & E,F,G,H         & Online        \\ 
\hline
\textbf{IBM Speech API}~\cite{ibm}               & Unknown             & \textcolor[rgb]{0.2,0.2,0.2}{Transcription} & Unknown                     & E,F,G,H         & Online        \\ 
\hline
Mozilla DeepSpeech~\cite{mozilla_ds} & RNN                 & \textcolor[rgb]{0.2,0.2,0.2}{Transcription} & End-to-End                  & E,F,G,H         & Local         \\ 
\hline
Intel Neon DeepSpeech~\cite{intel_ds}    & RNN                 & \textcolor[rgb]{0.2,0.2,0.2}{Transcription} & MFSC                        & I,J,K,L         & Local         \\ 
\hline
Kaldi~\cite{kaldi_hmm}              & HMM-GMM             & \textcolor[rgb]{0.2,0.2,0.2}{Transcription} & MFCC                        & E,F,G,H         & Local         \\ 
\hline
Kaldi-DNN~\cite{kaldi_dnn}          & DNN                 & \textcolor[rgb]{0.2,0.2,0.2}{Transcription} & MFCC                        & E,F,G,H         & Local         \\ 
\hline
Sphinx~\cite{sphinx}             & HMM-GMM             & \textcolor[rgb]{0.2,0.2,0.2}{Transcription} & MFCC                        & E,F             & Local         \\ 
\hline
\textbf{Wit.ai Speech API}~\cite{wit}               & Unknown             & \textcolor[rgb]{0.2,0.2,0.2}{Transcription} & Unknown                     & E,F,G,H         & Online        \\
\hline

\end{tabular}
\caption{The models we tested our perturbation attack scheme against. The Phrase ID is referenced from Table~\ref{tab:phrases_table}. The proprietary VPSes are in \textbf{bold}. }
\label{tab:model_table}
\vspace{-1em}
\end{table*}

\subsection{Attack Audio Generation}

\begin{figure} 
  \centering
    \includegraphics[width=0.5\textwidth]{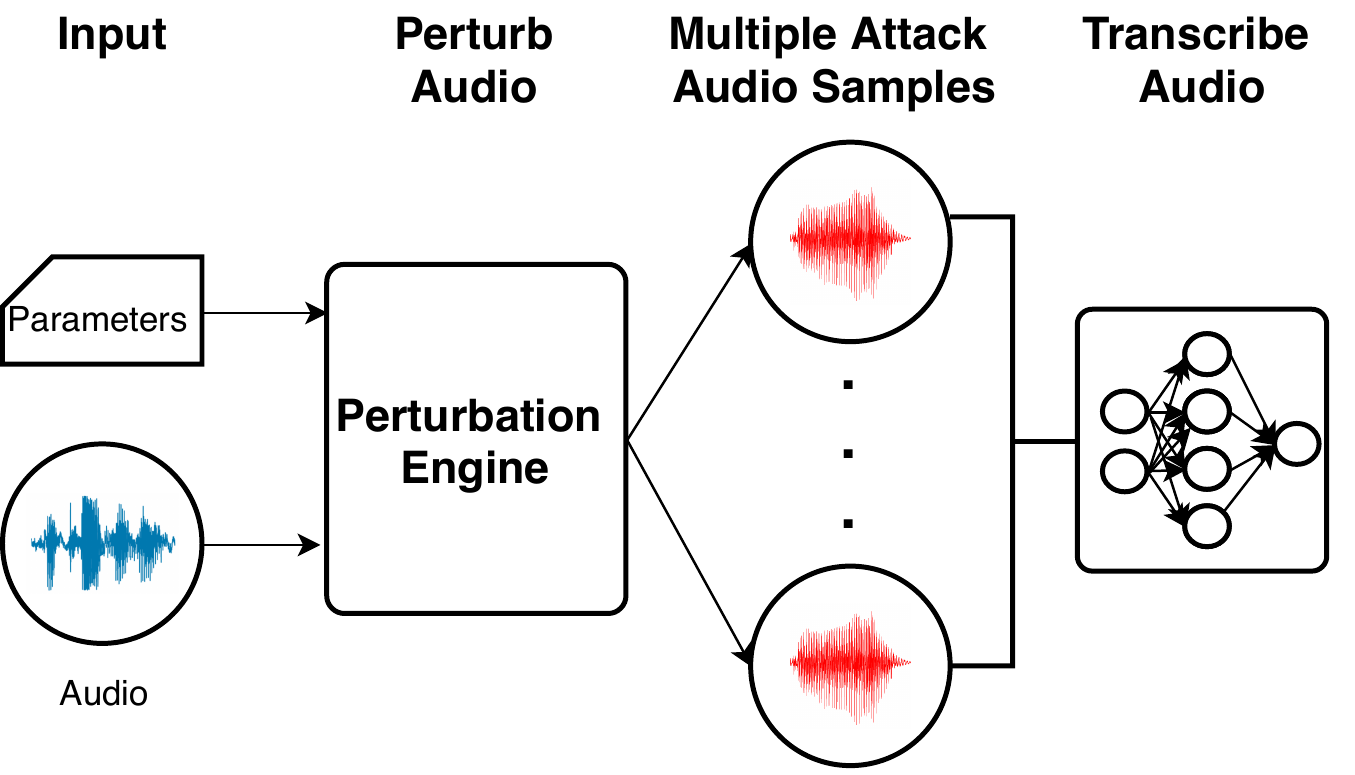}
  \caption{The workflow for our Attack Audio Generation. The Perturbation Engine takes as input the perturbation parameters and an audio file. It generates attack audio samples that that sound like noise to the human ear, but are transcribed correctly by the transcription model.}
\label{fig:architecture}
\vspace{-1em}
\end{figure}

This section details how the attack audio is generated using the
Perturbation Engine (PE), shown in Figure~\ref{fig:architecture}. The PE
works for any model given our black-box assumption.

\subsubsection{Generic Attack Method}\label{generic-attack}

Attack audio generation comprises of the following steps: parameter
selection, attack audio generation, model inference, and audio
selection. An overview of this section is illustrated in
Figure~\ref{fig:architecture}. 

First, the attacker selects the parameters with which to perturb the
audio: audio speed, high frequency intensity, and window size (for TDI and RPG). The attacker does not
know which parameters will give the most distorted attack audio the VPS
will still accept. The PE distorts the audio based on the input
parameters.

Second, the attacker feeds these parameters along with the normal audio
to the PE. The PE generates attack audio samples, perturbing the entire
length of the audio for each parameter set using the perturbations
described in the previous section. For each set of parameters, the PE
will generate a single audio sample. The TDI and
RPG perturbation schemes take the window size as
input. The HFA perturbation scheme will take as
input the frequency one wants to add to the audio signal. TS
perturbation scheme will take as input the percentage by which to increase
the tempo.

The PE will apply the schemes, using the designated input parameter
values, to generate a unique attack audio for each set of parameters.
Generating individual attack audio samples using our PE takes fractions
of a second.

Third, attack audio samples are passed to the target VPS, via queries.
The VPS will not accept all attack audio samples, as some might be
distorted beyond recognition. A trade-off exists between the degree of
distortion and model transcription. If the audio distortion is high, the
VPS might not accept or incorrectly transcribe the audio. On the other
hand, too little distortion will increase audio decipherability by a
human observer.  

\begin{figure} 
  \centering
    \includegraphics[width=0.5\textwidth]{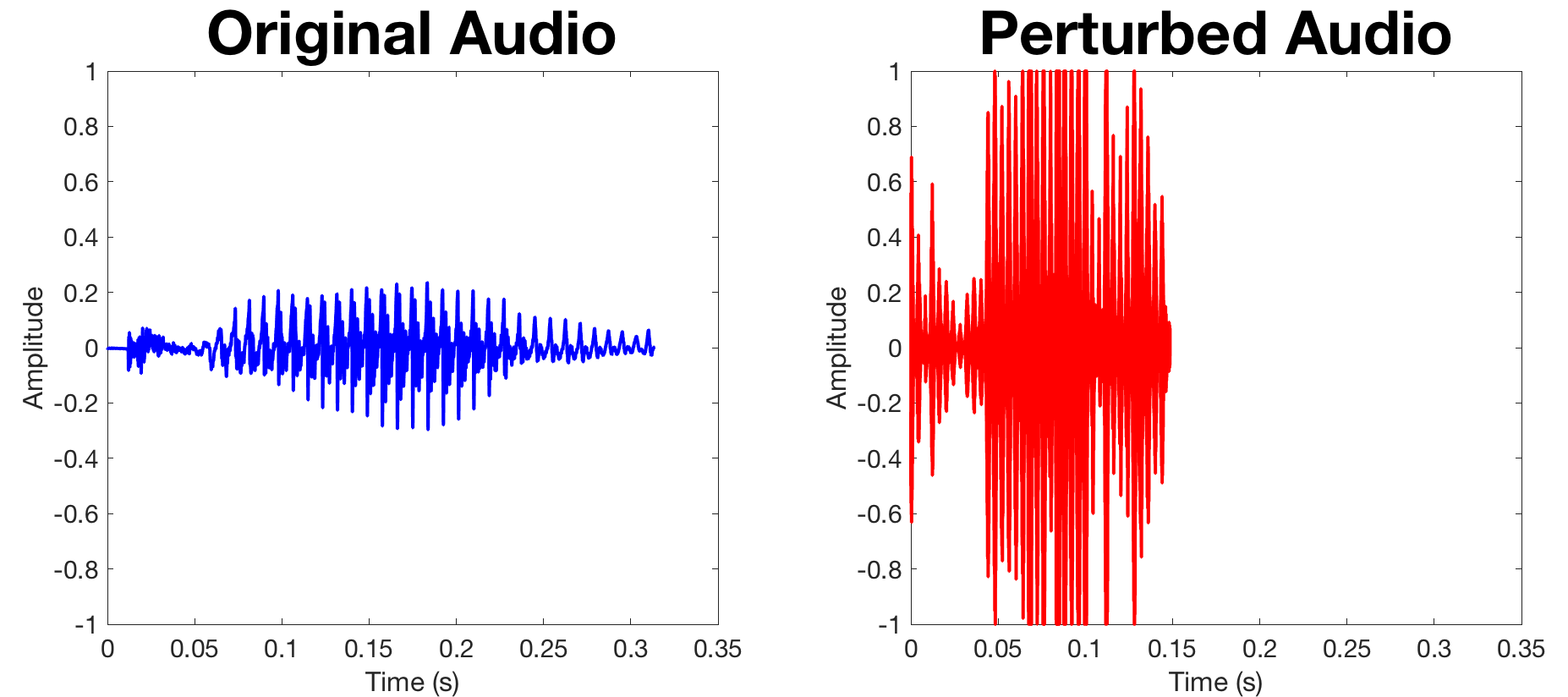}
  \caption{The audio recording of the word ``pay'' before and after it has been distorted by our perturbation techniques.}
\label{fig:perturbed_all}
\vspace{-1em}
\end{figure}

\subsubsection{Improved Attack Method}\label{improved-attack}

VPSes recognize some words better than others. At times, this property
is intentionally built into models. For example, home assistants are
specifically trained to be sensitive to activation phrases. Therefore,
some words can be degraded more than others and still be recognized by a
VPS.

The attacker can exploit this model bias towards certain words and
phrases by running the PE at a finer granularity, for each word rather
than the entire sentence. For example, rather than perturbing the entire
phrase ``pay money'', the words ``pay'' and ``money'' can be perturbed
individually as shown in Figure~\ref{fig:perturbed_all}. Next, the
perturbed audio samples for each word are concatenated to create all the
possible combinations, resulting in a much larger set of attack audio
samples.

\subsection{Over-the-Line and Over-the-Air}\label{over}

We define two scenarios to test the perturbation attacks against models:
\textit{Over-the-Line} and \textit{Over-the-Air}. In an Over-the-Line
attack the attack audio is passed to the model directly, as a
\texttt{.wav} file. Conversely, an \textit{Over-the-Air} attack requires
the adversary to play the attack audio via a speaker towards the target
VPS. It is not necessarily the case that an attack audio that is
successful Over-the-Line will also be successful Over-the-Air.  Playing
the attack audio via a speaker will degrade the attack audio's fine
perturbations due to natural factors that include acoustic hardware's
frequency response, static interference, and environmental noise. For a
practical attack, the attack audio should be resistant to these factors.

This limitation is evident in all state-of-the-art attacks against
VPSes. While Carlini et al.'s~\cite{carlini2016hidden} attack audio
cannot be used over any speaker systems other than the one used during
the attack generation process, Yuan et
al.'s~\cite{yuan2018commandersong} attack method assumes knowledge and
access to the attack speaker and the victim's acoustic hardware.
Although both cases present strong threat models in the real world, our
attacks have no such limitations, as we will demonstrate.

\subsection{Models}

Our proposed attack is aimed at the state of the art in publicly
available VPS implementations. We test our attack against the VPSes
listed in Table~\ref{tab:model_table}. Our selection captures a wide
range of known architectures, black-box APIs, and feature extraction
techniques. We give a brief overview of the different categories these
VPSes reside in and why they were chosen. Note that although the
architecture of some systems are known, only acoustic properties are
used to construct attack audio rather than any available information
about the underlying model. 

\subsubsection{Traditional Models}

We treat Kaldi's Hidden Markov Model-Gaussian Mixture Model (HMM-GMM)
combination as an example of a traditional speech transcription model.
Kaldi is well known due to its open-source nature, reasonable
performance, and flexible configurations. An HMM is a multi-state
statistical model which can serve as a temporal model in the context of
speech transcription. Each state in the HMM holds a unique GMM, which
takes MFCC-based features as input, and models phonemes in the context
of other phonemes. The HMM models state transitions and can account for
different clusters of temporal phoneme information to produce an
alignment between speech and text. HMM-based approaches are relevant
because modern architectures mimic their multi-state behavior. We refer
to ~\cite{Rabiner89atutorial} for a general primer on Hidden Markov
Models. 

\subsubsection{Modern Models}

More recently, speech recognition and transcription tasks have begun to employ
some combination of Deep Neural Networks (DNNs) and Recurrent Neural
Networks (RNNs). These architectures are popular due to their
performance and flexibility in complex automation tasks. Neural networks
consist of one or more hidden layers of stateful neurons sandwiched
between an input layer and an output layer. Layers are connected
together and act as inputs into subsequent layers, such that connections
are activated according to certain inputs. DNNs are neural networks with
more hidden layers and carefully designed deep architectures. \textcolor{black}{Convolutional Neural Networks (CNNs) are a popular DNN architecture due to their performance in image classification tasks, which derives from its ability to automatically perform feature extraction using convolution filters. RNNs differ slightly from DNNs, and can
be thought of as a chain of independent neural networks, such that
hidden layers of one neural network connect to the hidden layers of a
subsequent network. However, a basic RNN is susceptible to the problems of vanishing or exploding gradients~\cite{pmlr-v28-pascanu13}. Thus, RNNs are often implemented with special neuronal activation functions, such as Long Short Term Memory (LSTM) or Gated Recurrent Units (GRU). In either case, the inputs and outputs of an RNN may form sequences, making them ideal for many sequence-to-sequence tasks. These include language modeling and audio transcription, which are key components of modern ASRs. }

The selected neural network models differ in terms of feature
extraction. Kaldi-DNN and Intel Neon DeepSpeech use MFCC-based features
as input to the model. Until recently, MFCC-based features have been the
most powerful feature extraction technique for VPS-related tasks.
However, DeepSpeech-style models rely on an approach known as
\textit{end-to-end} learning. \textcolor{black}{In this system, both feature extraction and inference are
performed by the neural network model.
In the case of Mozilla DeepSpeech, the RNN is trained to translate raw audio spectrograms into word transcriptions. The goal of end-to-end audio transcription models is to learn a new feature space that will be more performant than traditional MFCC-based encodings. This feature space is used as input into the
model's RNN, which performs inference based on the encoded data. At the decoding stage, an optional language model may be used to improve transcription performance. Thus, neural networks are used between each end of the system. In practice,
this system exhibits state-of-the-art results for speech transcription
metrics, rivalling complex HMM-GMM methods}.

\subsubsection{Online APIs}

We validate our attack against several Internet-hosted models. The
architecture of these models is unknown, but we assume them to be near
state-of-the-art due to their popularity and widespread use. Of these
models, two are dedicated to speaker classification tasks: the Azure
Attestation API, is meant to classify a speaker among some known group,
while the Azure Verification API is meant to perform attestation based
on a user's voice. All other models are designed for speech
transcription tasks.

\section{Experimental Setup}\label{sec:setup}

\subsection{Phrase Selection}
The full list of phrases used for the experiments is shown in Table~\ref{tab:phrases_table}. For attacks against Speaker Identification models, we chose phrases A, B, C and D. Phrases A, B and C are phonetically balanced sentences from the TIMIT dataset~\cite{garofolo1988getting}. The phrases E, F, G, and H are command phrases used for attacks against ASRs. These represent commands an adversary would potentially want to use. Because these phrases were not interpretable by the Intel DeepSpeech model, we replaced them with phrases I, J, K and L. These additional phrases were collected from the LibriSpeech Dataset that the model was trained on~\cite{panayotov2015librispeech}. We did not include any activation phrases such as ``Ok Google", as every major VPS is tuned to detect its own activation phrase. This would mean we could perturb the activation phrase more than others, leading to biased results. 
\begin{table}[t]
\centering

\begin{tabular}{|l|c|l|c|}
\hline
\textbf{ID}        & \multicolumn{1}{l|}{\textbf{Model}} & \textbf{Phrase}                                                                        & \textbf{Success Rate (\%)} \\ \hline
\textbf{A} & \multirow{4}{*}{Identification}     & When suitably lighted                                                                  & 100                        \\ \cline{1-1} \cline{3-4} 
\textbf{B} &                                    & \begin{tabular}[c]{@{}l@{}}Don't ask me to carry \\ an oily rag like that\end{tabular} & 100                        \\ \cline{1-1} \cline{3-4} 
\textbf{C} &                                    & What would it look like                                                                & 100                        \\ \cline{1-1} \cline{3-4} 
\textbf{D} &                                    & \begin{tabular}[c]{@{}l@{}}My name is unknown \\ to you\end{tabular}                   & 100                        \\ \hline
\textbf{E} & \multirow{8}{*}{ASR}               & Pay money                                                                              & 100                        \\ \cline{1-1} \cline{3-4} 
\textbf{F} &                                    & Run browser                                                                            & 100                        \\ \cline{1-1} \cline{3-4} 
\textbf{G} &                                    & Open the door                                                                          & 100                        \\ \cline{1-1} \cline{3-4} 
\textbf{H} &                                    & Turn on the computer                                                                   & 100                        \\ \cline{1-1} \cline{3-4} 
\textbf{I} &                                    & Spinning indeed                                                                        & 100                        \\ \cline{1-1} \cline{3-4} 
\textbf{J} &                                    & Very Well                                                                              & 100                        \\ \cline{1-1} \cline{3-4} 
\textbf{K} &                                    & The university                                                                         & 100                        \\ \cline{1-1} \cline{3-4} 
\textbf{L} &                                    & Now to bed boy                                                                         & 100                        \\ \hline
\end{tabular}
\caption{We used sentences from the TIMIT corpus~\cite{garofolo1988getting}, which provides phonetically balanced sentences used widely in the audio testing community.}\label{tab:phrases_table}
\vspace{-1em}
\end{table}

\subsection{Speech Transcription} We tested our audio perturbation attack methodology against ten ASRs shown in Table~\ref{tab:model_table}. If our hypothesis is correct, the attack audio should work against any ASR, regardless of the underlying model type or feature extraction algorithm.

We ran the generic attack method (described in Section~\ref{generic-attack}), against proprietary models that include 7 proprietary (e.g., Google Speech API, Bing Speech API, IBM Speech API, Azure API etc)~\cite{uberi}. These models were hosted on-line and could only be queried a limited number of times. We ran the improved attack method (described in Section~\ref{improved-attack}), against locally hosted models, which could be queried without limits.
\subsubsection{Over-the-Line}
For each phrase, we generated multiple attack audio samples using various perturbation parameters. These were then passed to the model as \texttt{.wav} files for transcription. We then picked the single worst sounding audio as the final attack audio, based on criteria we describe Section~\ref{sec:discussion}. These steps were repeated for each model. At the end of the process, for each of the 10 models, we had one attack audio file for each phrase referenced in Table~\ref{tab:model_table}. 

\subsubsection{Over-the-Air} 
We ran the attack audio samples in Over-the-Air as described in Section~\ref{over}. Of the ten ASRs, we tested seven Over-the-Air. A single Audioengine A5 speaker~\cite{speaker} was placed on a conference room table to play the attack audio. A Behringer microphone~\cite{microphone} was placed one foot away to simulate an ASR device. The recorded audio was then passed to the ASR for transcription as a \texttt{.wav} file. For each model, we played the four attack audio samples, one for each phrase. The Mozilla DeepSpeech model was tested extensively by playing 15 randomly sampled attack audio files. This was done to ascertain whether audio samples with generated with larger window sizes to the PE still worked Over-the-Air. This was done as a baseline test to ensure that the attack algorithm did not inherently degrade the audio such that the attack audio ceased to remain effective after it was played Over-the-Air.

To show that the attack was independent of the acoustic hardware, we repeated the Over-the-Air experiment for the Google Client Speech API. Additionally, we degraded the acoustic environment with a white-noise generator playing at 55dB in the background. We replaced the previous speaker with an iMac and the microphone with the Motorola Nexus 6. The experiment occurred in a lab cubicle. This experimental setup represents harsh, yet realistic, acoustic conditions that an attacker faces in the modern world.

\subsection{Speaker Verification and Attestation}
Home assistant systems have begun to introduce voice biometrics to authenticate users to prevent VPSes from accepting audio commands from unauthorized individuals. Each audio command is checked against a voice blueprint of the real user to authenticate the source before it is carried out. This rudimentary biometrics authentication poses an obstacle which state-of-the-art audio obfuscation attacks can not overcome. However, our obfuscation technique is designed to retain the necessary voice information to pass the authentication tests. To test this hypothesis, we ran our audio obfuscation attack against two speaker recognition systems. We specifically attacked the Azure Identification/Attestation (SA) model and the Azure Verification (SV) model.

SA models are trained to identify the voices of multiple speakers. For example, during training a model learns to identify the voices of Alice and Bob. During testing, the model infers which speaker a test audio sample belongs to. Additionally, SA is text-independent. This means the audio sample can contain the voice recording of the speaker reading any text. 

In contrast, SV models are trained to identify a single speaker (i.e., just Bob). During testing, the model decides whether a test audio belongs to Bob or not. Unlike SA, SV is text-dependent as the test audio must contain the speaker reading a phrase requested by the model. The model first checks if the audio transcribes to the requested phrase. Only then does it check whether the voice blueprint matches that of Bob.

To attack an SA model, we first trained the Azure Identification model using the phrases from the TIMIT dataset. We trained the model on the voices of eight male and eight female speakers. Three male and three female speakers were randomly selected for the attack. For each speaker, we ran the perturbation scheme using the Generic Attack Method (described in Section ~\ref{generic-attack}) to generate 10 attack audio samples as \texttt{.wav}.

To attack an SV model, we trained the Azure Verification model. As the SV model is text-dependent, the TIMIT dataset was insufficient, as it does not contain any of the phrases that the Azure SV model requests. We gathered the audio data from three men and three women. We recorded four audio samples of each phrase per person: three samples to train the model and one sample to test. After training the model, we checked if the test audio authenticated correctly. The test audio was then perturbed using the Generic Attack Method and passed to the authentication model as \texttt{.wav} files. We repeated these steps individually for all six participants. 
\section{Experimental Results}\label{sec:results}

The Perturbation Engine takes as input a parameter set and an audio sample. It then generates attack audio samples according to the techniques outlined in Section~\ref{sec:types_perturb}. To demonstrate its effectiveness, we first test the Over-the-Line attack by providing the attack audio as a \texttt{.wav} file input to the VPS and the Over-the-Air attack by playing the audio via a commodity speaker.

\subsection{Over-the-Line}
\subsubsection{ASR Models}
We test our attack against ASR models. As noted previously, perturbations introduce acoustic artifacts that make the original message difficult for humans to understand. However, the attack audio must still transcribe correctly when passed to the ASR. 

We ran our attacks against ten ASRs shown in Table~\ref{tab:model_table}. The attack was successful if the ASR transcribed the entire phrase correctly. Minor spelling mistakes in the transcriptions were ignored. In each case, it took a few minutes to find the first attack audio that the ASR transcribed correctly. For each model we were able to produce a sample audio that the ASR correctly identified, as seen in Table~\ref{tab:phrases_table}, despite having no information about the ASR's underlying model or pre-processing steps. In total, we generated approximately 20,000 correctly transcribed attack audio samples. By design, modern ASRs are susceptible to replay attacks. Once an adversary generates an attack audio that works, they can use it repeatedly against multiple victims who use the same ASR. 

\begin{table}[t]
\centering
\begin{tabular}{|l|c|c|c|c|}
\hline
\textbf{Models}             & \textbf{\begin{tabular}[c]{@{}c@{}}Over \\ -the- \\Air\end{tabular}} &	\textbf{\begin{tabular}[c]{@{}c@{}}Attack \\ Type\end{tabular}} & \textbf{\begin{tabular}[c]{@{}c@{}}Min \\ TDI  \\Size (ms)\end{tabular}}	 & \textbf{\begin{tabular}[c]{@{}c@{}}Max TS \\Factor\\(\%)\end{tabular}} \\ \hline
Bing Speech API 			& 4/4               & TDI	& 3.36	 & -\\ \hline
Google Speech API           & 4/4               & TDI	& 1.47	 & -\\ \hline
Houndify 					& 3/4               & TDI	& 1.00	 & -\\ \hline
IBM Speech API 				& 3/4               & TDI	& 2.42	 & -\\ \hline
Mozilla DeepSpeech 			& 15/15             & TDI	   & 2.00 & -	\\ \hline
Kaldi-DNN 					& 4/4               & TDI+TS	& 1.00	 & 300\\ \hline
Wit.ai Speech API 			& 2/4               & TDI	& 1.94 	 & -\\ \hline
\end{tabular}
\caption{For phrases E-H in Table II, we generated 20 attack audio samples. Approximately 80\% of those samples successfully fooled the model for each of the 4 phrases. Manual listening tests revealed that the most obfuscated samples were generated using TDI and TS. These were then played Over-the-Air and were able to trick the model in nearly every case. Ambient noise is believed to have impacted the few non-passing phrases.}
\label{tab:over_air_table}
\vspace{-1em}
\end{table}

Almost all ASRs employ techniques that fall within the categories we covered in Table~\ref{tab:model_table}. We attack a diverse range of ASRs and all ASRs make similar assumptions about human speech. That is why it is reasonable to assume that our attack is effective against other state of the art ASRs, irrespective of the underlying mechanisms.

\begin{figure*} 
  \centering
    \includegraphics[width=0.8\textwidth]{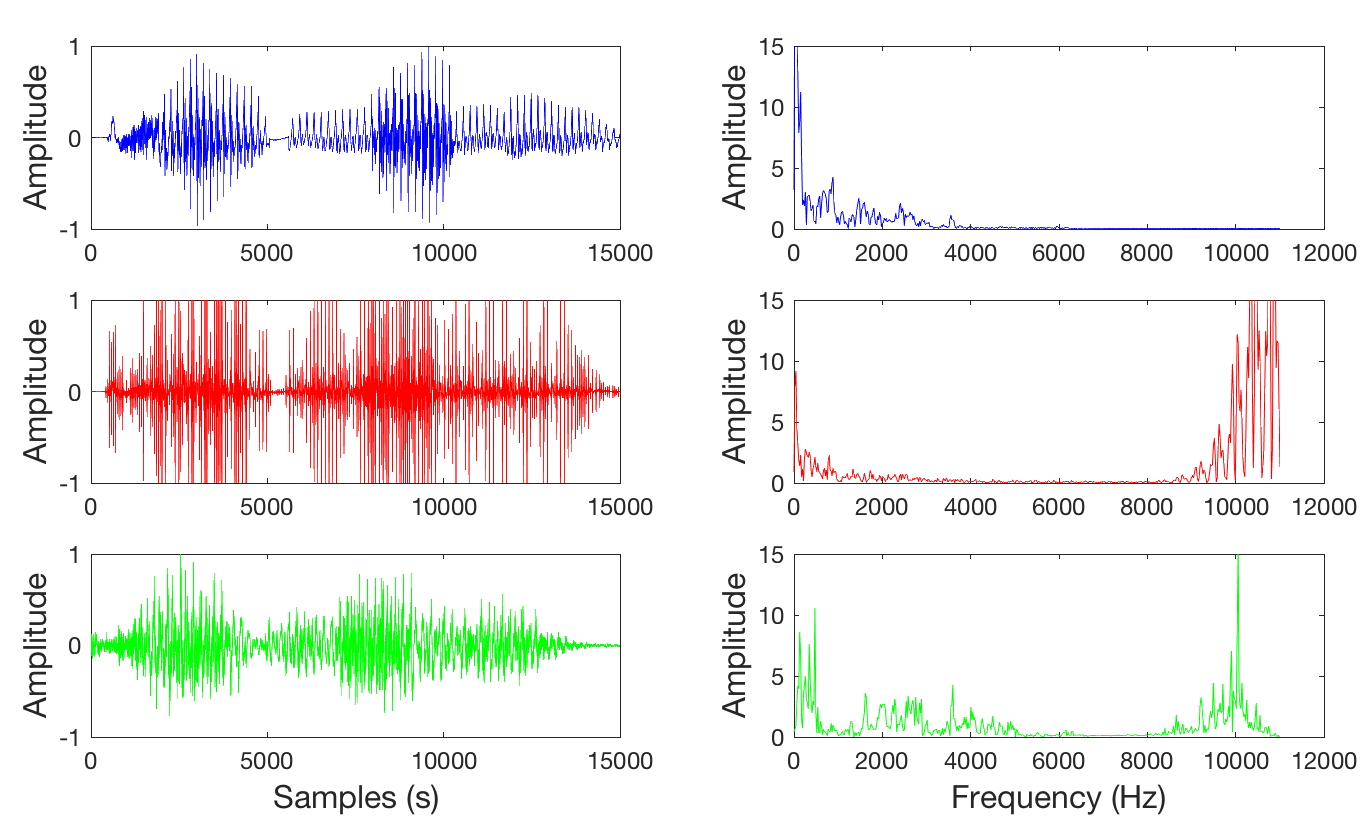}
  \caption{The time and frequency domain representation of the original audio sample (TOP), the perturbed audio after being played Over-the-Line (MIDDLE) and the perturbed audio after being played Over-the-Air (BOTTOM)}
\label{fig:three_comp}
\vspace{-1em}
\end{figure*}

\subsubsection{Speaker Identification Models}
Table~\ref{tab:model_table} shows the results for the attacks against speaker identification models. For this experiment, the attack was successful if the identification model classified the attack audio as that of the original speaker. In all cases, the models made the correct classification. The speaker models do not return any information about the audio's success, other than the classification. This means there is no information an attacker can rely on to improve the attack. Therefore, an offline perturbation scheme, like the one we propose, is preferable. 


\subsection{Over-the-Air}
In addition to being easy to generate, the attack audio must be correctly transcribed by the VPS after it has been played Over-the-Air. To test this against a randomly sampled set of seven of the ten ASRs, we use the same audio from the previous section. Figure~\ref{fig:three_comp} shows an example of the attack. The attack is a success if the model completely transcribed the desired phrase. The attack is unsuccessful if the ASR transcribes one or more words incorrectly. For example, if the word ``turn'' in the phrase ``turn on the computer'' was not transcribed, the attack is a failure. This is a conservative approach as an adversary can overcome this by using any of the 20,000 other attack audio samples that were created in the previous section.

\textcolor{black}{For the Over-the-air tests, we used the attack audio samples generated during the Over-the-Line experiments. For each of the four phrases E-H in Table~\ref{tab:phrases_table}, we picked the single most distorted audio sample. The attack parameters used to generate the audio samples are shown in Table~\ref{tab:over_air_table}. We used the TDI method, as this method generated the most audible distortion. For Kaldi-DNN, we performed TS in addition to TDI. Table~\ref{tab:over_air_table} shows the minimum window size for TDI. Any window size greater than this would lead to a successful attack Over-the-Air. Similarly, any TS factor smaller than the one in used Table~\ref{tab:over_air_table} led to a successful attack Over-the-Air.}

 As seen in Table~\ref{tab:over_air_table}, almost all phrases are successfully transcribed. This is expected as the attack is designed to retain the acoustic properties that a state of the art speaker model considers most important. The small drop in accuracy Over-the-Air, like in the case for Wit.ai, can be attributed to environmental factors (e.g. background noise etc).



Kaldi-DNN is the only model that did not transcribe any of the initial Over-the-Air attack audio samples successfully. Upon further inspection, we realized that the Kaldi-DNN model was trained using the Fisher English Dataset~\cite{cieri2004fisher}. We believe that the audio training data does not contain enough noisy audio. Therefore, when we introduce our own audio artifacts, the attack fails. However, when choosing different attack audio samples that are less perturbed, generated with larger window sizes as input to PE, we can show that the model transcribes them correctly. Failure of attack audio transcription does not mean that the model is immune to the perturbation attacks that we propose. Rather, it is possible the model is simply less effective in noisy environments.

\subsection{General Insights}
Our preliminary results revealed insights about parameter values that an attacker can use to reduce the attack generation time. Accordingly, we have demonstrated our hypothesis to hold for VPSes.

The experimental results show that perturbation parameters, specifically window size (used for TDI and RPG), display three important properties. First, the smaller the window size, the greater the audible distortion. Second, if an attack audio is successfully transcribed at a certain window size, then all attack audio samples that were generated with greater window sizes are also successfully transcribed. Third, no attack audio samples generated with window sizes of below of 1.00 ms are correctly transcribed. \textcolor{black}{This parameter creates the approximate upper bound for maximum distortion. Attack audio files generated with a smaller window size parameter did not transcribe correctly by any of the VPSes from Table~\ref{tab:model_table}.} Adversaries can use these three properties in conjunction to narrow down the number of attack audio samples they want to generate. Adversaries can start by setting the window size to 1.00 ms, when passing the perturbation parameters into the PE. They can increase the window size until they get the first attack audio that successfully transcribes.

Additionally, an audio sample can only be perturbed up to a certain threshold before it is distorted beyond VPS recognition. Attack audio samples that were generated with larger window sizes could be sped up the most. If an attacker perturbs audio by a single parameter's lowest bound (e.g., window size of 1.00 ms), using additional perturbation parameters, like speed, will reduce the likelihood of the audio being recognizable by the VPS. Attackers can choose different parameter settings to tune the attack audio to the victim and specific attack scenario to maximize attack efficacy. In our preliminary experiments we observed the TS factor of 150\%, RPG or TDI window size of near 1.00 ms, and \textcolor{black}{HFA of sine waves of} frequency above 8000 Hz produced the ideal attack audio.

\textcolor{black}{As discussed in Section~\ref{improved-attack}, the PE can be used to generate a large set of attack audio samples for a single phrase. Each audio sample has a different degree of audible distortion. However, picking the single worst attack sample is not straightforward, as there does not exist any widely accepted metric to measure the degree of perceived audio distortion. In our case, we identified the relationship between the perturbation parameters and audio distortion. For TDI and RPG, smaller window sizes correspond to greater distortion and thus worse sounding audio. For HFA and TS, larger values corresponded to worse sounding audio. These criteria allowed us to narrow the space of 20,000 attack audio samples, that we had generated previously, to less than ten. At this point, the attacker can manually listen to the ten samples, and pick the one that sounds the worst.}

Some perturbations were more successful than others at generating attack samples that the model could interpret. Of the perturbations that we tested, RPG was less successful than TDI. RPG requires us to take the FFT of the audio, followed by an inverse FFT. As we treat the models as black-boxes, we do not know the type or parameters (number of buckets) of the FFT algorithm that the model employs. In contrast, when using TDI, all the perturbations happen in the time domain. During this perturbation, the audio is not converted via a function like the FFT, though RPG did indeed work many times. With additional information about the model, we believe that the RPG mechanism could be tuned precisely; however, our approach demonstrates that blackbox knowledge is more than sufficient to launch successful attacks. 

\textcolor{black}{Additionally, the TDI perturbation does not require that the discontinuities align with the time frames of the feature extraction algorithm. This assumption is buttressed by the fact that perturbation schemes are successful against black-box models. In the case of these models, we do not have any information about the time frame discontinuities.}

When attacking an online model, an adversary is limited by the number of queries they can make. This could be either due to cost associated per each query or due to threat of being detected. That is why it is important that an adversary should be able to generate attack audio with the least number of queries possible. Our attack algorithm allows for exactly that. An adversary can use the RPG or TDI methods to generate ten attack audio samples, starting at the window size of 1.00 ms and using increments of 0.50 ms. We observe that it was almost always the case that at least one of the generated audio samples, is correctly interpreted by the VPS. In our case, we were able to successfully find an attack audio that worked for proprietary models in less than ten queries to the model. 

\begin{figure*}[t]
  \centering
    \includegraphics[width=0.75\linewidth]{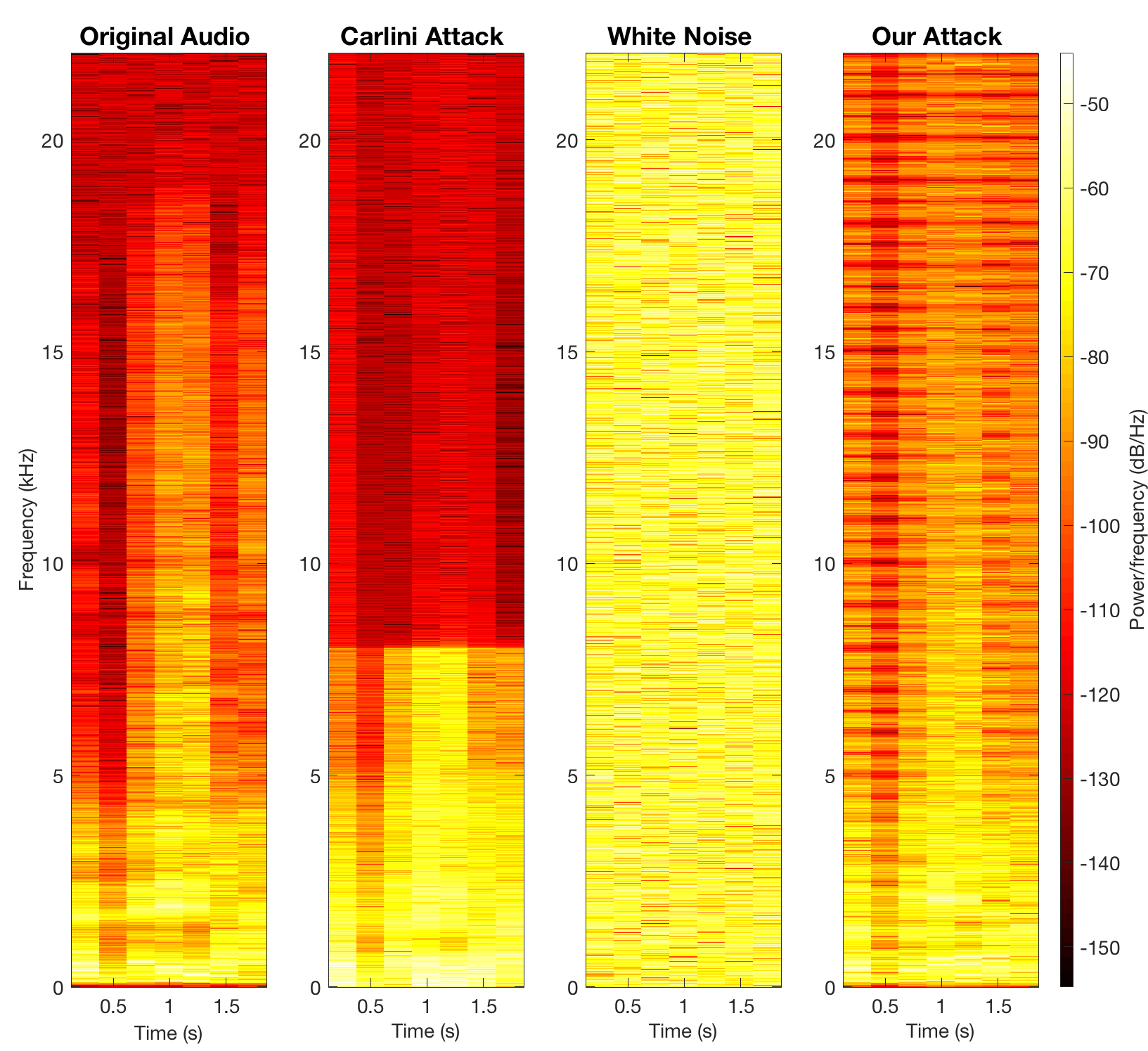}
    \caption{This figure displays an STFT of an audio file containing speech, Carlini's perturbation of the audio file, our perturbation of the audio file, and white noise. The lighter colors indicates a higher intensity at a given frequency and time. We can see that Carlini's perturbed audio has high intensities in lower frequencies and our perturbed audio has high intensities in the higher frequencies.}
\label{fig:attack_comp}
\vspace{-1em}
\end{figure*}

\textcolor{black}{This paper includes the largest evaluation of an attack in terms of number and variety of models attacked. Our model list exceeds that of any other published work in the area of adversarial audio by an order of magnitude. However, due to the evolving landscape of ML, new models will be released continuously. We show that the attack is sufficiently generalized by being effective on the existing, publicly available state-of-the-art models. The attack is `universal' in that it is designed to work against any model by only targeting the feature extraction phase. Intuitively, future models will be derived from the same feature extraction techniques available in the public domain. An additional obstacle lies in choosing a reasonable amount of phrases to test against. Running experiments against every phrase is intractable due to the amount of data ASRs are normally trained on. Instead we constructed a list of phrases that mirror the variety of real-world use cases.}


\section{Discussion}\label{sec:discussion}
\subsection{Improvements on Previous Work}
Our audio perturbation methods make attack audio samples difficult for humans to interpret due to psychoacoustic principles. Carlini et al.~\cite{carlini2016hidden} proposed a method with a similar goal to ours; however, their work adds background noise to the audio. Because humans can better recover speech from noise than our perturbations, our generated speech commands are comparatively harder for humans to interpret. We reinforce this claim both quantitatively and qualitatively.

\noindent\textbf{High frequencies}
Figure~\ref{fig:attack_comp} displays a Short-time Fourier Transform (STFT) of original audio sample, white noise, and both our and Carlini's attack on this original audio sample.\footnote{We are grateful to these authors, who willingly provided us with a small number of attack and original samples} The STFT plot provides information about the intensity of frequencies over time for each audio sample. The higher the intensity, the lighter the color. By looking at these plots we can see that the audio produced by Carlini et al. has greater intensities in the lower frequencies than in the higher frequencies. In contrast, our attack audio has greater intensities across the spectrum. 

Humans perceive loudness of audio on a logarithmic scale~\cite{fletcher1933loudness}, and higher frequencies are \textit{perceived} as being louder~\cite{ISO226}. This increase in perceived relative volume will lead listeners to concentrate more on high frequency components to the exclusion of low frequency components, which are normally associated with speech. Increasing the volume may also result in reaching the listener's pain tolerance~\cite{TED}. If the volume of the audio is decreased in order to hear the higher frequencies at a comfortable level, the lower frequencies become harder to hear.

The addition of higher frequencies in our attack audio also reduces intelligibility due to the Cocktail Party Effect~\cite{cherry,CPE1,CPE2}. This psychoacoustic phenomenon allows us to focus on a conversation while ignoring other conversation and background noise. By the similar concept of selective attention~\cite{select-1}, humans are more likely to properly interpret speech that is more familiar to them (such as a friendly voice). These familiar stimuli can cause a listener to switch their attention to interpreting other speech.

Though normally seen as robust qualities of human hearing, our attack audio capitalizes on these principles. The dominant high frequencies in our attacks are above are in the typical range of human speech~\cite{TED}. When heard by a human, our attack audio will most likely be considered background noise, and thus ignored. Additionally, treating the audio as unimportant background noise will reduce the chance that the speech will be a familiar stimuli and trigger a listener's attention. The high frequencies that exploit these psychoacoustic principles will be filtered out during the preprocessing stage of a VPS and will still result in an accurate transcription.

\noindent\textbf{Noise}
Though adding noise may seem like a reasonable method of increasing the unintelligibility of audio, studies in psychoacoustics have demonstrated the robustness of speech in the presence of noise~\cite{noise}. When noise is added to otherwise normal speech, the
auditory properties of speech stand out above the added noise thus ensuring intelligibility. If enough noise is added to speech, interpretation will become more difficult, because original speech becomes suppressed by the large amount of noise. Adding noise to speech as a perturbation method will have little benefit in terms of making the audio harder for humans to interpret. Additionally, as additional noise is added to the attack audio, it also becomes difficult for a VPS to accurately transcribe the audio, due to the suppression of the speech.

To evaluate their attack audio intelligibility, Carlini et al. used Amazon Mechanical Turk~\cite{turk}, asking participants to listen to their attack audio and attempt to transcribe it. Though this may seem like a suitable evaluation method, this presents many uncontrolled variables that could have a significant impact on their results. The issue with using Amazon Turk is that the experiment was not conducted locally, resulting in many uncontrollable variables that could affect a participant's ability to transcribe their attack audio. These variables include, but are not limited to,
age~\cite{select-1, age}, first language~\cite{fam}, range of hearing~\cite{TED}, listening environment, audio equipment, and visual stimuli~\cite{visual_stim}.

Quantifying the intelligibility of speech is challenging~\cite{difficult_quant}, but these survey results have too many uncontrollable variables to make an accurate conclusion. \textcolor{black}{ There does not exist any widely accepted metric to measure unintelligibility of an audio sample. The $L_2$ norm has been used in previous works, discussed in Section~\ref{sec:relwork}, to quantify the distortion of adversarial images when attacking image classifiers. However, to use the $L_2$ norm to measure attack audio distortion would be incorrect. This is because} we base the unintelligibility of our audio on principles of psychoacoustics and the biology of human hearing which is substantially and fundamentally different from factors associated with image classification. We have made our attack audio available to the public online.\footnote{Our attack: https://sites.google.com/view/practicalhiddenvoice } Carlini et al. also make their audio available online\footnote{Carlini et al.'s attack: http://www.hiddenvoicecommands.com/black-box},which we encourage the reader use as a comparison.

Performing a small-scale study, rather than
using the established results from another scientific community, is
redundant and prone to error; in effect, we would be merely reporting
anecdotes. Instead, we validate our results by citing
widely accepted published work from psychoacoustics which shows that our
audio has the properties that make it unintelligible. As Figure~\ref{fig:attack_comp} clearly demonstrates, our audio better exhibits the characteristics associated with unintelligibility than the previous work.

\noindent\textbf{Practical Limitations}
Carlini's attack is only possible under the assumption that the adversary has white-box knowledge of the victim's model. The attack can only be carried out against HMM-GMM ASRs, meaning that the attack is insufficient against state of the art DNN and RNN models, which are increasingly being deployed. The attack has not been shown to work against speaker identification models either. Carlini's attack audio can not be used with any speaker system other than the one used during the attack audio generation. Lastly, their white-box attack takes an upwards of 32 hours to generate. 
These additional factors severely limit the adversary's capabilities. In contrast, our attack is black-box, model-agnostic, effective against both ASRs and speaker identification models, transferable across speaker systems and takes seconds to generate. Our attack algorithm provides greater capabilities to the attacker, while simultaneously making fewer assumptions about the victim model. \textcolor{black}{ Lastly, our work exploits the fundamental nature of human audio perception to generate attack samples. Our perturbations make use of simple operations, in the time domain, without using any complex algorithms. In contrast, Carlini's work attempts to add noise to the MFCCs, which exist in the frequency domain. The domains both attacks exploit are fundamentally different, which impacts the effectiveness of Carlini's attack.}

\subsection{Defenses}
\noindent\textbf{Voice Activity Detection (VAD)}
VAD is a commonly implemented speech processing algorithm that is used to detect the presence of human voices in samples of audio. It has applications in cellular communications~\cite{diss_vad_tel}, Voice-over-IP (VoIP)~\cite{diss_vad_voip} and VPSes~\cite{diss_vad_asr}. VAD differentiates between the regions of human speech and regions of silence or noise in audio. By identifying silence and noise, those regions can be eliminated from the audio. This includes the regions of noise and silence between words, reducing the audio being transmitted to just individual words that make up the entire recording. This is particularly useful for VPSes as using VAD in preprocessing results in giving a model only the necessary speech data, potentially reducing processing cost and improving transcription results. 

Based on this description, VAD may be suggested as a potential defense against the attacks we present in this paper. If a VPS is using VAD, this would cause the attack audio to be classified as noise and \textit{not} be further processed. As we will demonstrate, this is not the case.

To test this potential defensive mechanism, we ran a VAD algorithm on a set of 36 randomly selected attack audio files. We executed a MATLAB implementation of the ITU-T G.729 VAD algorithm on the audio files and observed which regions were classified as speech. In all 36 cases, the VAD algorithm accurately located the speech in the audio.

In addition, we also ran the VAD algorithm over two\footnote{The VAD experiment requires both an attack audio file and the original source speech file. We could only run the VAD experiment for what was available to us of Carlini{'s} attack files.} attack audio files produced by Carlini et al. For both files almost all of the audio, including sections between the words consisting of pure noise, was determined to be speech. We believe this is the result of the added noise suppressing the speech, and in turn, ``blurring" the lines between speech and noise. While this does not prevent their attack audio from being given to an VPS, this does increase the chance that the audio will not be transcribed correctly. The noise that is considered to be speech, especially that between the words, will be sent along with the actual speech. This increases the chance that some or all of the speech could be mistranslated, preventing the attack.

As more systems start to deploy VAD to reduce processing time and server load, the attack audio produced by Carlini et al. may not continue to be transcribed correctly. Meanwhile, our attack audio will continue to work whether or not a system uses VAD for preprocessing because we did not introduce additional noise.

\noindent\textbf{Classifier Training}
The defensive method of training a classifier model to detect adversarial attacks on VPSes has previously been suggested~\cite{carlini2016hidden}. We believe this technique to be impractical and brittle. Machine learning models are imperfect regardless of training or tasks they are given. Adding a model to preemptively determine if audio is malicious would most likely result in usability issues. Legitimate speech commands may be incorrectly classified as malicious and ignored because of imperfections in the model. This would decrease the quality of the user experience and could potentially result in users looking to alternative VPS systems. 

\noindent\textbf{Altering the VPS}
Another potential defense mechanism would be to modify the construction of the VPS in order to hinder our attack. This could consist of steps such as altering or removing signal processing or preprocessing steps. However, tailoring a VPS to not detect or discard the attack audio is not a viable solution. This will result in a decrease in accuracy of the VPS. Additionally, we have demonstrated the effectiveness of our attack audio against Mozilla DeepSpeech, which does not use any predefined signal processing. Due to our perturbations being rooted in psychoacoustics, in order to impede our attack, a defense mechanism would have to be placed at or before the preprocessing steps.  

Such a system has been recently proposed~\cite{logan} and could potentially have an effect on the success of our attack. The idea behind this system is that audio produced by an electro-acoustic transducer (loudspeaker) will contain low frequencies in the sub-bass region (20-60 HZ). This is below the speaking range of most humans and is a good indicator that the speech was played by a loudspeaker. If this were to be implemented during preprocessing, this could identify our attack as being produced by a non-human speaker and prevent it from being given to the VPS. \textcolor{black}{Similarly, ASRs that use liveness detection~\cite{zhang2017hearing}~\cite{zhang2016voicelive} will also be relatively robust against our attack. However, using such defenses to identify malicious commands might generate false-positives, which degrade user experience and product popularity.}

\subsection{Limitations}
\textcolor{black}{The attack audio will have reduced effectiveness when used in noisy environments. This is because noise interferes with the attack audio making it difficult for VPSs to interpret. However, legitimate commands \textit{also} do not work well in noisy environments. Readers who have attempted to use their voice assistants in a noisy environment have likely experienced such a degradation.}

\section{Related Work}
\label{sec:relwork}

From personal assistant systems to speaker identification and
investments, machine learning (ML) is becoming increasingly incorporated
into our daily lives. Unfortunately, ML models are inherently vulnerable
to a wide spectrum of attacks.
Early research focused on enhancing train-time resilience for scenarios
where the adversary is capable of poisoning training data (e.g., spam
and worm detection)~\cite{dalvi2004, newsome2006}.  These early works
exhibited attack vectors that were later formalized into three axes: i)
influence of the adversary (causative, train-time attacks vs.
exploratory, test-time attacks), ii) type of security violation
(integrity vs. denial of service), and iii) attack specificity (targeted vs.
indiscriminate)~\cite{Barreno2006, Barreno2010}. A detailed overview of
each axis is provided by Huang et al.~\cite{Huang2011}. Although all
axes provide utility to an adversary, recent attacks have focused on a
narrow subsection of these axes.  

Following this categorization, much of the work in adversarial machine
learning has focused on \textit{exploratory targeted} attacks against
image classifiers in terms of both \textit{availability} and
\textit{integrity}. These attacks vary from changing particular
pixels~\cite{kurakin2016adversarial, szegedy2013intriguing,
goodfellow2014explaining, baluja2017adversarial, su2017one,
moosavi2016deepfool} or patches of pixels~\cite{brown2017adversarial,
sharif2016accessorize, papernot2016limitations, carlini2017towards} to
creating entirely new images that will classify to a chosen
target~\cite{nguyen2015deep,liu2017trojaning}.  Although these attacks are very
successful against image models, they do not suffice for attacking audio
models. Modern image models operate directly on the supplied image
pixels to derive relevant spatial
features~\cite{papernot2016limitations, nguyen2015deep}. Audio models,
in contrast, do not operate on individual samples, and instead derive
representations of the original temporal space using acoustic properties
of the human voice. This layer of temporal feature extraction adds a
layer of complexity to potential attacks, which means that small changes
to individual samples will likely never propagate to the final feature
space used for inference. 

Current attacks against audio models can be broken down into three broad
categories. The first involves generating malicious audio commands that
are completely inaudible to the human ear but are recognized by the
audio model~\cite{zhang2017dolphinattack}. The second embeds malicious
commands into piece of legitimate audio (e.g., a
song)~\cite{yuan2018commandersong}. The third obfuscates an audio
command to such a degree that the casual human observer would think of
the audio as mere noise but would be correctly interpreted by the victim
audio model~\cite{vaidya2015cocaine, carlini2016hidden}.  Our work falls
within the third category and closest attack to ours is that of Carlini
et al.~\cite{carlini2016hidden}. The success of this earlier work is
limited due to the following reasons: i) the attack can only be using
against an Hidden Markov Model-Gaussian Mixture Model architecture; ii)
the attack assumes the attacker has white-box access to the model; and
iii) the attack is slow, taking at least 32 hours of execution.  Our
attack is designed to overcome these limitations.

\section{Conclusion}
\label{sec:conclusion}

The security community has recently discovered a number of weaknesses in
specific machine learning models that underpin multiple VPSes. Such
weaknesses allow an attacker to inject commands that are unintelligible
to humans, but are still transcribed correctly by VPSes. While effective against
particular models using particular speakers and microphones, these
previous attack are not acoustic hardware independent or widely practical. Instead of
attacking underlying machine learning models, this paper instead
investigates generating attack audio based on the feature vectors
created by signal processing algorithms. With this black-box approach,
we demonstrate experimentally that our techniques work against a wide
array of both speech detection and speaker recognition systems both
Over-the-Wire and Over-the-Air. Moreover, because we treat
psychoacoustics as a principal element of our design, we are able to
explain why our attacks are less intelligible than prior work. In so
doing, we not only argue that hidden command attacks are practical, but
also that securing such systems must therefore take greater domain
knowledge of audio processing into consideration.

\section{Acknowledgment}
We would like to thank our reviewers, in particular our shepherd Dr. Jay Stokes, for their insightful comments and suggestions. We would also like to thank the authors of the Hidden Voice Commands paper~\cite{carlini2016hidden} who graciously provided us with both original and attack audio samples. Our comparison would not have been possible without their assistance. This work was supported in
part by the National Science Foundation under grant number
CNS-1526718 and 1540217. Any opinions, findings, and conclusions or
recommendations expressed in this material are those of the
authors and do not necessarily reflect the views of the National
Science Foundation. 

\bibliographystyle{IEEEtranS}
\bibliography{biometrics}

\begin{thebibliography}{10}
\providecommand{\url}[1]{#1}
\csname url@samestyle\endcsname
\providecommand{\newblock}{\relax}
\providecommand{\bibinfo}[2]{#2}
\providecommand{\BIBentrySTDinterwordspacing}{\spaceskip=0pt\relax}
\providecommand{\BIBentryALTinterwordstretchfactor}{4}
\providecommand{\BIBentryALTinterwordspacing}{\spaceskip=\fontdimen2\font plus
\BIBentryALTinterwordstretchfactor\fontdimen3\font minus
  \fontdimen4\font\relax}
\providecommand{\BIBforeignlanguage}[2]{{%
\expandafter\ifx\csname l@#1\endcsname\relax
\typeout{** WARNING: IEEEtranS.bst: No hyphenation pattern has been}%
\typeout{** loaded for the language `#1'. Using the pattern for}%
\typeout{** the default language instead.}%
\else
\language=\csname l@#1\endcsname
\fi
#2}}
\providecommand{\BIBdecl}{\relax}
\BIBdecl

\bibitem{turk}
\BIBentryALTinterwordspacing
``Amazon mechanical turk,'' \url{https://www.mturk.com/}. [Online]. Available:
  \url{https://www.mturk.com/}
\BIBentrySTDinterwordspacing

\bibitem{speaker}
``Audioengine a5+,''
  \url{https://audioengineusa.com/shop/poweredspeakers/a5-plus-powered-speakers/}.

\bibitem{azure_attest}
``Azure identification api,''
  \url{https://github.com/Microsoft/Cognitive-SpeakerRecognition-Python/tree/master/Identification}.

\bibitem{azure_verify}
``Azure verification api,''
  \url{https://github.com/Microsoft/Cognitive-SpeakerRecognition-Python/tree/master/Verification}.

\bibitem{microphone}
``Behringer,'' \url{https://www.musictri.be/brand/behringer/home}.

\bibitem{bing}
``Bing speech api,''
  \url{https://azure.microsoft.com/en-us/services/cognitive-services/speech/}.

\bibitem{google}
``Cloud speech-to-text,'' \url{https://cloud.google.com/speech-to-text/}.

\bibitem{houndify}
``Houndify,'' \url{https://www.houndify.com/}.

\bibitem{ibm}
``Ibm speech to text,''
  \url{https://www.ibm.com/watson/services/speech-to-text/}.

\bibitem{intel_ds}
``Intel implementation of deep speech 2 in neon,''
  \url{https://github.com/NervanaSystems/deepspeech}.

\bibitem{kaldi_dnn}
``Kaldi aspire chain model,'' \url{http://kaldi-asr.org/models.html}.

\bibitem{mozilla_ds}
``Mozilla project deepspeech,''
  \url{https://azure.microsoft.com/en-us/services/cognitive-services/speaker-recognition/}.

\bibitem{uberi}
``Uberi speech recognition modules for python,''
  \url{https://github.com/Uberi/speech_recognition}.

\bibitem{wit}
``Wit.ai natural language for developers,'' \url{https://wit.ai/}.

\bibitem{ISO226}
\BIBentryALTinterwordspacing
``{ISO} 226:2003,'' \url{https://www.iso.org/standard/34222.html}, 2003.
  [Online]. Available: \url{https://www.iso.org/standard/34222.html}
\BIBentrySTDinterwordspacing

\bibitem{amodei2016deep}
D.~Amodei, S.~Ananthanarayanan, R.~Anubhai, J.~Bai, E.~Battenberg, C.~Case,
  J.~Casper, B.~Catanzaro, Q.~Cheng, G.~Chen \emph{et~al.}, ``Deep speech 2:
  End-to-end speech recognition in english and mandarin,'' in
  \emph{International Conference on Machine Learning}, 2016, pp. 173--182.

\bibitem{Angluin1992}
\BIBentryALTinterwordspacing
D.~Angluin, ``Computational learning theory: Survey and selected
  bibliography,'' in \emph{Proceedings of the Twenty-fourth Annual ACM
  Symposium on Theory of Computing}, ser. STOC '92.\hskip 1em plus 0.5em minus
  0.4em\relax New York, NY, USA: ACM, 1992, pp. 351--369. [Online]. Available:
  \url{http://doi.acm.org/10.1145/129712.129746}
\BIBentrySTDinterwordspacing

\bibitem{baluja2017adversarial}
S.~Baluja and I.~Fischer, ``Adversarial transformation networks: Learning to
  generate adversarial examples,'' \emph{arXiv preprint arXiv:1703.09387},
  2017.

\bibitem{Barreno2010}
M.~Barreno, B.~Nelson, A.~D. Joseph, and J.~D. Tygar, ``{The security of
  machine learning},'' \emph{Machine Learning}, vol.~81, no.~2, pp. 121--148,
  2010.

\bibitem{Barreno2006}
\BIBentryALTinterwordspacing
M.~Barreno, B.~Nelson, R.~Sears, A.~D. Joseph, and J.~D. Tygar, ``{Can machine
  learning be secure?}'' in \emph{Proceedings of the 2006 ACM Symposium on
  Information, computer and communications security - ASIACCS '06}, 2006,
  p.~16. [Online]. Available:
  \url{https://www.cs.drexel.edu/{~}greenie/cs680/asiaccs06.pdf
  http://portal.acm.org/citation.cfm?doid=1128817.1128824}
\BIBentrySTDinterwordspacing

\bibitem{logan}
L.~Blue, L.~Vargas, and P.~Traynor, ``Hello, is it me you`re looking for?
  differentiating between human and electronic speakers for voice interface
  security,'' in \emph{11th ACM Conference on Security and Privacy in Wireless
  and Mobile Networks}, 2018.

\bibitem{CPE2}
A.~W. Bronkhorst, ``The cocktail-party problem revisited: early processing and
  selection of multi-talker speech,'' \emph{Attention, Perception, \&
  Psychophysics}, vol.~77, no.~5, pp. 1465--1487, 2015.

\bibitem{brown2017adversarial}
T.~B. Brown, D.~Man{\'e}, A.~Roy, M.~Abadi, and J.~Gilmer, ``Adversarial
  patch,'' \emph{arXiv preprint arXiv:1712.09665}, 2017.

\bibitem{carlini2016hidden}
N.~Carlini, P.~Mishra, T.~Vaidya, Y.~Zhang, M.~Sherr, C.~Shields, D.~Wagner,
  and W.~Zhou, ``Hidden voice commands.'' in \emph{USENIX Security Symposium},
  2016, pp. 513--530.

\bibitem{carlini2017towards}
N.~Carlini and D.~Wagner, ``Towards evaluating the robustness of neural
  networks,'' in \emph{Security and Privacy (SP), 2017 IEEE Symposium
  on}.\hskip 1em plus 0.5em minus 0.4em\relax IEEE, 2017, pp. 39--57.

\bibitem{cherry}
E.~C. Cherry, ``Some experiments on the recognition of speech, with one and
  with two ears,'' in \emph{The Journal of the Acoustical Society of America,
  25th}.\hskip 1em plus 0.5em minus 0.4em\relax
  \url{http://www.ee.columbia.edu/~dpwe/papers/Cherry53-cpe.pdf}, 1953, p.
  975–979.

\bibitem{cieri2004fisher}
C.~Cieri, D.~Miller, and K.~Walker, ``The fisher corpus: a resource for the
  next generations of speech-to-text.'' in \emph{LREC}, vol.~4, 2004, pp.
  69--71.

\bibitem{dalvi2004}
\BIBentryALTinterwordspacing
N.~Dalvi, P.~Domingos, Mausam, S.~Sanghai, and D.~Verma, ``{Adversarial
  classification},'' \emph{Proceedings of the 2004 ACM SIGKDD international
  conference on Knowledge discovery and data mining - KDD '04}, p.~99, 2004.
  [Online]. Available:
  \url{http://portal.acm.org/citation.cfm?doid=1014052.1014066}
\BIBentrySTDinterwordspacing

\bibitem{noise}
\BIBentryALTinterwordspacing
R.~L. Diehl, ``Acoustic and auditory phonetics: the adaptive design of speech
  sound systems,'' \emph{Philosophical Transactions of the Royal Society B:
  Biological Sciences}, vol. 363, p. 965–978, 2008. [Online]. Available:
  \url{/url{https://www.ncbi.nlm.nih.gov/pmc/articles/PMC2606790/pdf/rstb20072153.pdf}}
\BIBentrySTDinterwordspacing

\bibitem{fletcher1933loudness}
H.~Fletcher and W.~A. Munson, ``Loudness, its definition, measurement and
  calculation,'' \emph{Bell System Technical Journal}, vol.~12, no.~4, pp.
  377--430, 1933.

\bibitem{garofolo1988getting}
J.~S. Garofolo \emph{et~al.}, ``Getting started with the darpa timit cd-rom: An
  acoustic phonetic continuous speech database,'' \emph{National Institute of
  Standards and Technology (NIST), Gaithersburgh, MD}, vol. 107, p.~16, 1988.

\bibitem{masking_book}
S.~A. Gelfand, \emph{Hearing: An Introduction to Psychological and
  Physiological Acoustics}, 5th~ed.\hskip 1em plus 0.5em minus 0.4em\relax
  Informa Healthcare, 2009.

\bibitem{CPE1}
\BIBentryALTinterwordspacing
S.~Getzmann, J.~Jasny, and M.~Falkenstein, ``Switching of auditory attention in
  “cocktail-party” listening: Erp evidence of cueing effects in younger and
  older adults,'' \emph{Brain and Cognition}, vol. 111, pp. 1 -- 12, 2017.
  [Online]. Available:
  \url{http://www.sciencedirect.com/science/article/pii/S0278262616302408}
\BIBentrySTDinterwordspacing

\bibitem{goodfellow2014explaining}
I.~J. Goodfellow, J.~Shlens, and C.~Szegedy, ``Explaining and harnessing
  adversarial examples,'' \emph{arXiv preprint arXiv:1412.6572}, 2014.

\bibitem{graves2014towards}
A.~Graves and N.~Jaitly, ``Towards end-to-end speech recognition with recurrent
  neural networks,'' in \emph{International Conference on Machine Learning},
  2014, pp. 1764--1772.

\bibitem{hanley1949effect}
T.~D. Hanley and G.~Draegert, ``Effect of level of distracting noise upon
  speaking rate, duration, and intensity.'' PURDUE RESEARCH FOUNDATION
  LAFAYETTE IND, Tech. Rep., 1949.

\bibitem{hannun2014deep}
A.~Hannun, C.~Case, J.~Casper, B.~Catanzaro, G.~Diamos, E.~Elsen, R.~Prenger,
  S.~Satheesh, S.~Sengupta, A.~Coates \emph{et~al.}, ``Deep speech: Scaling up
  end-to-end speech recognition,'' \emph{arXiv preprint arXiv:1412.5567}, 2014.

\bibitem{diss_vad_voip}
B.~Hartpence, \emph{Packet Guide to Voice over IP}, 1st~ed.\hskip 1em plus
  0.5em minus 0.4em\relax O`Reilly Media, Inc, 2013.

\bibitem{Huang2011}
\BIBentryALTinterwordspacing
L.~Huang, A.~D. Joseph, B.~Nelson, B.~I. Rubinstein, and J.~D. Tygar,
  ``Adversarial machine learning,'' in \emph{Proceedings of the 4th ACM
  Workshop on Security and Artificial Intelligence}, ser. AISec '11.\hskip 1em
  plus 0.5em minus 0.4em\relax New York, NY, USA: ACM, 2011, pp. 43--58.
  [Online]. Available: \url{http://doi.acm.org/10.1145/2046684.2046692}
\BIBentrySTDinterwordspacing

\bibitem{kurakin2016adversarial}
A.~Kurakin, I.~Goodfellow, and S.~Bengio, ``Adversarial examples in the
  physical world,'' \emph{arXiv preprint arXiv:1607.02533}, 2016.

\bibitem{sphinx}
P.~Lamere, P.~Kwok, W.~Walker, E.~Gouvea, R.~Singh, B.~Raj, and P.~Wolf,
  ``Design of the cmu sphinx-4 decoder,'' in \emph{Eighth European Conference
  on Speech Communication and Technology}, 2003.

\bibitem{method_noise}
B.~P. Lathi and Z.~Ding, \emph{Modern Digital and Analog Communication
  Systems}, 4th~ed.\hskip 1em plus 0.5em minus 0.4em\relax Oxford University
  Press, 2009.

\bibitem{TED}
\BIBentryALTinterwordspacing
C.~Limb, ``Building the musical muscle,'' \emph{TEDMED}, 2011. [Online].
  Available:
  \url{\url={https://www.ted.com/talks/charles_limb_building_the_musical_muscle#t-367224}}
\BIBentrySTDinterwordspacing

\bibitem{liu2017trojaning}
Y.~Liu, S.~Ma, Y.~Aafer, W.-C. Lee, J.~Zhai, W.~Wang, and X.~Zhang, ``Trojaning
  attack on neural networks,'' \emph{Proceedings of the 2017 Network and
  Distributed System Security Symposium (NDSS)}, 2017.

\bibitem{BKAd}
S.~Maheshwari, ``{Burger King `O.K. Google' Ad Doesn't Seem O.K. With
  Google},''
  \url{https://www.nytimes.com/2017/04/12/business/burger-king-tv-ad-google-home.html},
  2017.

\bibitem{mel_scale}
\BIBentryALTinterwordspacing
R.~Mannell, ``The perceptual and auditory implications of parametric scaling in
  synthetic speech,'' \emph{Macquarie University}, p. (Chapter 2), 1994.
  [Online]. Available:
  \url{\url{http://clas.mq.edu.au/speech/acoustics/auditory_representations/pitchdiscrim.html}}
\BIBentrySTDinterwordspacing

\bibitem{CM17}
C.~Martin, ``{72\% Want Voice Control In Smart-Home Products},'' Media Post
  --\url{https://www.mediapost.com/publications/article/292253/72-want-voice-control-in-smart-home-products.html?edition=99353},
  2017.

\bibitem{moosavi2016deepfool}
S.~M. Moosavi~Dezfooli, A.~Fawzi, and P.~Frossard, ``Deepfool: a simple and
  accurate method to fool deep neural networks,'' in \emph{Proceedings of 2016
  IEEE Conference on Computer Vision and Pattern Recognition (CVPR)}, no.
  EPFL-CONF-218057, 2016.

\bibitem{newsome2006}
J.~Newsome, B.~Karp, and D.~Song, ``Paragraph: Thwarting signature learning by
  training maliciously,'' in \emph{Recent Advances in Intrusion Detection},
  D.~Zamboni and C.~Kruegel, Eds.\hskip 1em plus 0.5em minus 0.4em\relax
  Berlin, Heidelberg: Springer Berlin Heidelberg, 2006, pp. 81--105.

\bibitem{diss_vad_tel}
H.~Newton and S.~Schoen, \emph{Newton's Telecom Dictionary}, 30th~ed.\hskip 1em
  plus 0.5em minus 0.4em\relax Harry Newton, 2016.

\bibitem{nguyen2015deep}
A.~Nguyen, J.~Yosinski, and J.~Clune, ``Deep neural networks are easily fooled:
  High confidence predictions for unrecognizable images,'' in \emph{Proceedings
  of the IEEE Conference on Computer Vision and Pattern Recognition}, 2015, pp.
  427--436.

\bibitem{UKdollhouse}
S.~Nichols, ``{TV anchor says live on-air `Alexa, order me a dollhouse'- Guess
  what happens next},''
  \url{https://www.theregister.co.uk/2017/01/07/tv-anchor-says-alexa-buy-me-a-dollhouse-and-she-does/},
  2017.

\bibitem{age}
W.~G. on~Speech~Understanding and Aging, ``Speech understanding and aging,''
  \emph{The Journal of the Acoustical Society of America}, vol.~83, no.~3, pp.
  859--895, 1988.

\bibitem{panayotov2015librispeech}
V.~Panayotov, G.~Chen, D.~Povey, and S.~Khudanpur, ``Librispeech: an asr corpus
  based on public domain audio books,'' in \emph{Acoustics, Speech and Signal
  Processing (ICASSP), 2015 IEEE International Conference on}.\hskip 1em plus
  0.5em minus 0.4em\relax IEEE, 2015, pp. 5206--5210.

\bibitem{papernot2016limitations}
N.~Papernot, P.~McDaniel, S.~Jha, M.~Fredrikson, Z.~B. Celik, and A.~Swami,
  ``The limitations of deep learning in adversarial settings,'' in
  \emph{Security and Privacy (EuroS\&P), 2016 IEEE European Symposium
  on}.\hskip 1em plus 0.5em minus 0.4em\relax IEEE, 2016, pp. 372--387.

\bibitem{pmlr-v28-pascanu13}
\BIBentryALTinterwordspacing
R.~Pascanu, T.~Mikolov, and Y.~Bengio, ``On the difficulty of training
  recurrent neural networks,'' in \emph{Proceedings of the 30th International
  Conference on Machine Learning}, ser. Proceedings of Machine Learning
  Research, S.~Dasgupta and D.~McAllester, Eds., vol.~28, no.~3.\hskip 1em plus
  0.5em minus 0.4em\relax Atlanta, Georgia, USA: PMLR, 17--19 Jun 2013, pp.
  1310--1318. [Online]. Available:
  \url{http://proceedings.mlr.press/v28/pascanu13.html}
\BIBentrySTDinterwordspacing

\bibitem{select-1}
\BIBentryALTinterwordspacing
D.~J. Plude, J.~T. Enns, and D.~Brodeur, ``The development of selective
  attention: A life-span overview,'' \emph{Acta Psychologica}, vol.~86, no.~2,
  pp. 227 -- 272, 1994. [Online]. Available:
  \url{http://www.sciencedirect.com/science/article/pii/0001691894900043}
\BIBentrySTDinterwordspacing

\bibitem{kaldi_hmm}
D.~Povey, A.~Ghoshal, G.~Boulianne, L.~Burget, O.~Glembek, N.~Goel,
  M.~Hannemann, P.~Motlicek, Y.~Qian, P.~Schwarz, J.~Silovsky, G.~Stemmer, and
  K.~Vesely, ``The kaldi speech recognition toolkit,'' in \emph{IEEE 2011
  Workshop on Automatic Speech Recognition and Understanding}.\hskip 1em plus
  0.5em minus 0.4em\relax IEEE Signal Processing Society, 2011, iEEE Catalog
  No.: CFP11SRW-USB.

\bibitem{fam}
\BIBentryALTinterwordspacing
F.~Pulvermuller and Y.~Shtyrov, ``Language outside the focus of attention: The
  mismatch negativity as a tool for studying higher cognitive processes,''
  \emph{Progress in Neurobiology}, vol.~79, no.~1, pp. 49 -- 71, 2006.
  [Online]. Available:
  \url{http://www.sciencedirect.com/science/article/pii/S0301008206000323}
\BIBentrySTDinterwordspacing

\bibitem{Rabiner89atutorial}
L.~R. Rabiner, ``A tutorial on hidden markov models and selected applications
  in speech recognition,'' in \emph{PROCEEDINGS OF THE IEEE}, 1989, pp.
  257--286.

\bibitem{rabiner1978digital}
L.~R. Rabiner and R.~W. Schafer, \emph{Digital processing of speech
  signals}.\hskip 1em plus 0.5em minus 0.4em\relax Prentice Hall, 1978.

\bibitem{diss_vad_asr}
J.~Ram{\'i}rez, J.~M. G{\'o}rriz, and J.~C. Segura, ``Voice activity detection.
  fundamentals and speech recognition system robustness,'' in \emph{Robust
  speech recognition and understanding}, 2007.

\bibitem{sharif2016accessorize}
M.~Sharif, S.~Bhagavatula, L.~Bauer, and M.~K. Reiter, ``Accessorize to a
  crime: Real and stealthy attacks on state-of-the-art face recognition,'' in
  \emph{Proceedings of the 2016 ACM SIGSAC Conference on Computer and
  Communications Security}.\hskip 1em plus 0.5em minus 0.4em\relax ACM, 2016,
  pp. 1528--1540.

\bibitem{visual_stim}
\BIBentryALTinterwordspacing
B.~G. Shinn-Cunningham, ``Object-based auditory and visual attention,'' in
  \emph{Trends in Cognitive Science}, 2008, pp. 182--186. [Online]. Available:
  \url{\url{http://www.cns.bu.edu/~shinn/resources/pdfs/2008/2008TICS_Shinn.pdf}}
\BIBentrySTDinterwordspacing

\bibitem{HS18}
H.~Stephenson, ``{UX design trends 2018: from voice interfaces to a need to not
  trick people},'' {Digital Arts -
  }\url{https://www.digitalartsonline.co.uk/features/interactive-design/ux-design-trends-2018-from-voice-interfaces-need-not-trick-people/},
  2018.

\bibitem{difficult_quant}
\BIBentryALTinterwordspacing
A.~Stojanow and J.~Liebetrau, ``A review on conventional psychoacoustic
  evaluation tools, methods and algorithms,'' in \emph{2016 Eighth
  International Conference on Quality of Multimedia Experience (QoMEX)}, June
  2016, pp. 1--6. [Online]. Available:
  \url{https://ieeexplore.ieee.org/document/7498923/}
\BIBentrySTDinterwordspacing

\bibitem{su2017one}
J.~Su, D.~V. Vargas, and S.~Kouichi, ``One pixel attack for fooling deep neural
  networks,'' \emph{arXiv preprint arXiv:1710.08864}, 2017.

\bibitem{szegedy2013intriguing}
C.~Szegedy, W.~Zaremba, I.~Sutskever, J.~Bruna, D.~Erhan, I.~Goodfellow, and
  R.~Fergus, ``Intriguing properties of neural networks,'' \emph{arXiv preprint
  arXiv:1312.6199}, 2013.

\bibitem{torrey2010transfer}
L.~Torrey and J.~Shavlik, ``Transfer learning,'' in \emph{Handbook of Research
  on Machine Learning Applications and Trends: Algorithms, Methods, and
  Techniques}.\hskip 1em plus 0.5em minus 0.4em\relax IGI Global, 2010, pp.
  242--264.

\bibitem{vaidya2015cocaine}
T.~Vaidya, Y.~Zhang, M.~Sherr, and C.~Shields, ``Cocaine noodles: exploiting
  the gap between human and machine speech recognition,'' \emph{WOOT}, vol.~15,
  pp. 10--11, 2015.

\bibitem{venugopalan2014translating}
S.~Venugopalan, H.~Xu, J.~Donahue, M.~Rohrbach, R.~Mooney, and K.~Saenko,
  ``Translating videos to natural language using deep recurrent neural
  networks,'' \emph{arXiv preprint arXiv:1412.4729}, 2014.

\bibitem{yuan2018commandersong}
X.~Yuan, Y.~Chen, Y.~Zhao, Y.~Long, X.~Liu, K.~Chen, S.~Zhang, H.~Huang,
  X.~Wang, and C.~A. Gunter, ``Commandersong: A systematic approach for
  practical adversarial voice recognition,'' in \emph{{Proceedings of the
  USENIX Security Symposium}}, 2018.

\bibitem{zhang2017dolphinattack}
G.~Zhang, C.~Yan, X.~Ji, T.~Zhang, T.~Zhang, and W.~Xu, ``Dolphinattack:
  Inaudible voice commands,'' in \emph{Proceedings of the 2017 ACM SIGSAC
  Conference on Computer and Communications Security}.\hskip 1em plus 0.5em
  minus 0.4em\relax ACM, 2017, pp. 103--117.

\bibitem{zhang2017hearing}
L.~Zhang, S.~Tan, and J.~Yang, ``Hearing your voice is not enough: An
  articulatory gesture based liveness detection for voice authentication,'' in
  \emph{Proceedings of the 2017 ACM SIGSAC Conference on Computer and
  Communications Security}.\hskip 1em plus 0.5em minus 0.4em\relax ACM, 2017,
  pp. 57--71.

\bibitem{zhang2016voicelive}
L.~Zhang, S.~Tan, J.~Yang, and Y.~Chen, ``Voicelive: A phoneme localization
  based liveness detection for voice authentication on smartphones,'' in
  \emph{Proceedings of the 2016 ACM SIGSAC Conference on Computer and
  Communications Security}.\hskip 1em plus 0.5em minus 0.4em\relax ACM, 2016,
  pp. 1080--1091.

\end{thebibliography}

\end{document}